\documentclass[showpacs,preprintnumbers,amsmath,amssymb,latexsym,letterpaper]{revtex4}
\usepackage{graphicx}
\usepackage{setspace} 
\newcommand{\rem}[1]{#1}
\usepackage{dcolumn} 
\usepackage{bm}      
\usepackage{color} 
\usepackage{aas_macros}

\newcommand{\dofa}{{\sl dof} }
\newcommand{\dof}{{\sl dof}}

\newcommand{\leraya}{{Leray$-\alpha$} }
\newcommand{\clarka}{{Clark$-\alpha$} }
\newcommand{\lansa}{{LANS$-\alpha$} }
\newcommand{\leray}{Leray$-\alpha$}
\newcommand{\clark}{Clark$-\alpha$}
\newcommand{\lans}{LANS$-\alpha$}

\renewcommand{\vec}{\mathbf}

\def\sunu#1{#1}          
\def\add#1{#1}          
\def\thesis#1{}                          

\begin{document}

\title{Three regularization models of the Navier-Stokes equations}

\author{Jonathan {Pietarila Graham}$^{1,2}$, Darryl
D. Holm$^{3,4}$, Pablo D. Mininni$^{1,5}$, \\and Annick Pouquet$^{1}$}
\affiliation{$^{1}$National Center for Atmospheric Research, P.O. Box 3000, Boulder, Colorado 80307, USA\footnote{The National Center for Atmospheric Research is sponsored by the National Science Foundation} \\$^2${currently at Max-Planck-Institut f\"ur 
Sonnensystemforschung, 37191 Katlenburg-Lindau, Germany} \\ $^3$ Department of Mathematics, Imperial College London, London SW7 2AZ, UK \\ $^4$ Computer and Computational Science Division, Los Alamos National Laboratory , Los Alamos, NM 87545, USA \\ $^5$ Departamento de F\'\i sica, Facultad de Ciencias Exactas y Naturales, Universidad de Buenos Aires, Ciudad Universitaria, 1428 Buenos Aires, Argentina.}

\date{\small\today}


\begin{abstract}

\add{We determine how the differences in the treatment of the
sub-filter-scale physics affect the properties of the flow for three
closely related regularizations of Navier-Stokes.  The consequences on
the applicability of the regularizations as sub-grid-scale (SGS)
models are also shown by examining their effects on super-filter-scale
properties.  Numerical solutions of the Clark$-\alpha$ model are
compared to two previously employed regularizations, the
Lagrangian-Averaged Navier-Stokes $\alpha-$model (LANS$-\alpha$) and
Leray$-\alpha$ albeit at significantly higher Reynolds number than
previous studies, namely $Re
\approx 3300$, Taylor Reynolds number of $R_\lambda \approx790$,} and to a direct numerical simulation (DNS) of the
Navier-Stokes equations.  We derive the K\'arm\'an-Howarth equation
for both the Clark$-\alpha$ and Leray$-\alpha$ models.  We confirm one
of two possible scalings resulting from this equation for
Clark$-\alpha$ as well as its associated $k^{-1}$ energy spectrum.
\add{At sub-filter scales, Clark$-\alpha$ possesses similar total
dissipation and characteristic time to reach a statistical turbulent steady-state as
Navier-Stokes, but exhibits greater intermittency.  As a SGS model,}
Clark$-\alpha$ reproduces the large-scale energy spectrum and
intermittency properties of the DNS.  For the Leray$-\alpha$ model,
increasing the filter width, $\alpha$, decreases the nonlinearity and,
hence, the effective Reynolds number is substantially
decreased. Therefore even for the smallest value of $\alpha$ studied
Leray$-\alpha$ was inadequate as a SGS model. The LANS$-\alpha$ energy
spectrum $\sim k^1$, consistent with its so-called ``rigid bodies,''
precludes a reproduction of the large-scale energy spectrum of the DNS
\add{at high $Re$ while achieving a large reduction in numerical
resolution}.  We find, however, that this same feature reduces its
intermittency compared to Clark$-\alpha$ (which shares a similar
K\'arm\'an-Howarth equation).  Clark$-\alpha$ is found to be the best
approximation for reproducing the total dissipation rate and the
energy spectrum at scales larger than $\alpha$, whereas {high-order}
intermittency properties for larger values of $\alpha$ are best
reproduced by LANS$-\alpha$.
\end{abstract}


\maketitle 
\section{Introduction}

Nonlinearities prevail in {fluid dynamics} when the Reynolds number,
$Re$, is large \cite{F95}.  For geophysical flows, the Reynolds number
is often larger than $10^8$ and for some astrophysical flows values of 
$Re\approx 10^{18}$ is not unreasonable.  The number of degrees of freedom
(\dof) in the flow increases as $Re^{9/4}$ for $Re \gg 1$ in the
Kolmogorov framework \cite{K41a,K41b,K41c} (hereafter K41).  Such a huge number of \dofa makes direct numerical simulations (DNS) of  turbulence at high $Re$
infeasible on any existing or projected computer for decades to come.
Because of this intractability, simulations of  turbulence are always carried out in regions of parameter space far from the observed values, either
with: (a) an unphysical lack of scale separation between the
energy-containing, inertial, and dissipative ranges while 
parameterizing the missing physics, or (b) a study of the processes at
much smaller length scales, often with periodic boundaries
(unphysical at large scales but 
used under the hypothesis of homogeneity of turbulent flows).
Clearly, modeling of unresolved small scales is necessary.  

\add{Given the nonlinear nature of turbulent flows and the ensuing multi-scale interactions, the physics of the
unresolvable scales may not be separable from the properties (e.g.,
statistics) of the resolvable large scales.  However,} two main approaches have
been developed over the years to model the effects of the unresolvable
small scales in turbulence on the scales resolved in the simulations.
The first approach is Large Eddy Simulations (LES, see \cite{MK00}).
LES is widely used in engineering, in atmospheric sciences, and to a
lesser extent in astrophysics.  However, in the LES approach, the
Reynolds number is not known.  Instead, one attempts modeling the
behavior of the flow in the limit of very large $Re$.  \add{As the
Kolmogorov assumption of self-similarity is known to be violated
(e.g., by intermittency \cite{KIY+03,T04} and by spectral non-locality
\cite{AMP05b}), the value of $Re$ can play an important role, e.g., in
the competition between two or more instabilities \cite{MiPoMo2006}.
Therefore,} another approach models the effects of turbulence at
higher Reynolds numbers than are possible with a DNS on a given grid,
by using a variety of techniques that can be viewed as filtering of
the small scales \add{(the so-called sub-grid-scale (SGS) models).}

A rather novel approach to modeling of turbulent flows employs {\it
regularization modeling} as a SGS model
\cite{CHM+99,GH02b,GH03,HN03,CHO+05,GH06}.  Unlike closures which
employ eddy-viscosity concepts (modifying the dissipative processes),
the approach of regularization modeling modifies the spectral
distribution of energy.  \add{For this reason, they retain a
well-defined Reynolds number.}  Existence and uniqueness of smooth
solutions can be rigorously proven, unlike many LES models (e.g.,
eddy-viscosity), as well as the fact that the subgrid model recovers
the Navier-Stokes equations in the limit of the filter width going to
zero.  Their robust analytical properties ensure computability of
solutions.  \add{These same properties reopen theoretical
possibilities first explored by Leray when he} proved the existence
(but not smoothness, or uniqueness) of solutions to the Navier-Stokes
equations in $\mathbb R^n$, n=2,3 using the Leray model \cite{L34}.
\add{This treatment of the small scales, then, enforces a precise
type of regularization of the entire solution which may be studied as
an independent scientific question (as compared to either LES or SGS
modeling).}


Geurts and Holm \cite{GH02b,GH03,GH06} began using the Leray model with (a three-point invertible approximation of) an inverse-Helmholtz-operator filter of width $\alpha$.  Later it was dubbed \leraya and an upper bound for the
dimension of the global attractor was established \cite{CHO+05}.  The
global existence and uniqueness of strong solutions for the Leray
model is a classical result \cite{L34}.  \leraya has been compared to
 DNS simulations on a grid of $N^3=192^3$ in a doubly-periodic compressible channel flow domain
\cite{GH02b,GH03,GH06}.  Its performance was found to be superior to a dynamic mixed (similarity plus
eddy-viscosity) model (with an even greater reduction in computational
cost). However, it possessed a systematic error of a slight
over-prediction of the large scales accompanied by a slight
under-prediction of the small scales.  It possessed both forward-
and back-scatter, but exhibited too little dissipation.

The Leonard tensor-diffusivity model \cite{L74} (sometimes known as
the Clark model \cite{CFR79}) is the first term of the reconstruction
series for the turbulent sub-filter stress for all symmetric filters
that possess a finite nonzero second moment.  This leading order
approximation of the subgrid stress is thus generic
\cite{WWV+01,CWJ01}.  In {\sl a priori} testing, it reconstructs a
significant fraction ($>90\%$, but not all) of the subgrid stress,
provides for local backscatter along the stretching directions while
remaining globally dissipative, and possesses a better reconstruction
of the subgrid stress than the scale-similarity model.  {Used as a
LES,} in {\sl a posteriori} testing the Leonard tensor-diffusivity
model required additional dissipation (a dynamic Smagorinsky term) to
achieve reasonable gains in computation speed for 3D periodic flows
and for channel flows \cite{WWV+01}. The Leonard tensor-diffusivity
model {does not conserve energy in the non-viscous limit}.  Cao, Holm,
and Titi \cite{CHT05} developed a related (conservative) subgrid model
which they dubbed \clark.  The \clark\, model applies an additional
inverse-Helmholtz filter operation to the Reynolds stress tensor of
the Clark model. The global well-posedness of the \clarka model and
the existence and uniqueness of its solutions were demonstrated, and
upper bounds for the Hausdorff ($d_H$) and fractal ($d_F$) dimensions
of the global attractor were found \cite{CHT05}.  \add{This model has
yet to be evaluated numerically.}

The third regularization model we will consider is the incompressible
Lagrangian-averaged Navier-Stokes (\lans, $\alpha-$model, also known
as the viscous Camassa-Holm equation
\cite{HMR98b,CFH+98,CFH+99a,CHM+99,CFH+99b}).  It can be derived, for
instance, by applying temporal averaging to Hamilton's principle,
where Taylor's frozen-in turbulence hypothesis (the only approximation
in the derivation) is applied as the closure for the Eulerian
fluctuation velocity in the Reynold decomposition, at linear order in
the generalized Lagrangian mean description \cite{HMR98a,H02a,H02b}.
In this derivation, the momentum-conservation structure of the
equations is retained.  For scales smaller than the filter width,
\lansa reduces the steepness in gradients of the Lagrangian mean
velocity and thereby limits how thin the vortex tubes may become as
they are transported, while the effect on larger length scales is
negligible \cite{CHM+99}.  \lansa may also be derived by smoothing the
transport velocity of a material loop in Kelvin's circulation theorem
\cite{FHT01}.  Consequently, there is no attenuation of resolved
circulation, which is important for many engineering and geophysical
flows where accurate prediction of circulation is highly desirable.
An alternative interpretation of the $\alpha-$model is that it
neglects fluctuations in the smoothed velocity field, while preserving
them in the source term, the vorticity \cite{MP02}.

\lansa has previously been compared to direct numerical simulations
(DNS) of the Navier-Stokes equations at modest Taylor Reynolds numbers
($R_\lambda \approx 72$ \cite{ZM04}, $R_\lambda \approx 130$
\cite{CHM+99}, and $R_\lambda\approx 300$ \cite{CHO+05}).  \lansa was
compared to a dynamic eddy-viscosity LES in 3D isotropic turbulence
under two different forcing functions (for $R_\lambda \approx 80$ and
$115$) and for decaying turbulence with initial conditions peaked at a
low wavenumber (with $R_\lambda \approx 70$) as well as at a moderate
wavenumber (with $R_\lambda \approx 220$) \cite{MKS+03}.  \lansa was
preferable in these comparisons because it demonstrated correct
alignment between eigenvectors of the subgrid stress tensor and the
eigenvectors of the resolved stress tensor and vorticity vector.
\add{The LES effectiveness of} the \lansa and the \leraya
\add{regularization models relative to eddy-viscosity and the} dynamic
mixed model (similarity plus eddy-viscosity) \add{have already been
demonstrated} in a turbulent mixing shear layer (with $Re \approx 50$)
\cite{GH02a,GH06}.  \lansa was found to be the most accurate of these
three LES candidates at proper subgrid resolution, but the effects of
numerical contamination can be strong enough to lose most of this
potential.  While \lansa has the greatest grid-independent accuracy of
the three models, it also requires the greatest resolution.  From the
LES perspective, this could pose some limitations on the practical use
and application of \lansa for high $Re$ cases. {Indeed, recent
high-resolution simulations of \lansa showed that energy artificially
accumulates in the sub-filter-scales, giving as a result only a modest
computational gain at very high Reynolds number \cite{PGHM+07a}.}

We propose to pursue these previous studies of \leraya and \lansa
further at high\add{er} Reynolds number, \add{and to use them as a benchmark
for evaluation of \clark.}  One goal is to \add{contrast the sub-filter-scale physics of the three
models to determine the relevant features from which to build improved
models.}  As the three
regularizations are related via truncation of sub-filter stresses,
\add{such a comparison can be illuminating.}  For \lans, the predicted
\add{sub-filter}-scale spectra is $\sim k^{-1}$ \cite{FHT01}.  This scaling has
been observed to be subdominant to an energy spectrum $\sim k^1$ which
corresponds to ``enslaved rigid body'' or ``polymerized'' portions of
the fluid {\cite{PGHM+07a}}.  The sub-filter scaling observed in the
third-order structure function corresponded to the predicted $\sim
k^{-1}$ scaling of the energy spectrum. However, {regions were
observed in the flow where no stretching was acting in the sub-filter
scales. These regions, which give no contribution to the energy
cascade, and hence do not affect the third order structure functions,
are responsible for the $\sim k^1$ scaling in the \lansa energy
spectrum.}  For \clark, the correct time scale for vortex stretching
is difficult to determine and its spectrum is found to range between
$\sim k^{-1}$ and $\sim k^{-7/3}$ {\cite{CHT05}}.  \leraya has the
same difficulty and the spectrum can range between $\sim k^{-1/3}$ and
$\sim k^{-5/3}$ {\cite{CHO+05}}.  The determination of these scaling
laws is needed to quantify the computational gain if each model is to
be used as a SGS model.  As a result, \add{we seek} to determine
empirically the sub-filter scale spectra.  \add{Our second goal is to
evaluate the applicability of these three regularizations as SGS
models.  This is accomplished both through prediction of computational
gains from observed sub-filter-scale properties and through directly
testing their capability to predict super-filter-scale properties at
high $Re$.}

We present the three models and describe how they are related, derive
the K\'arm\'an-Howarth equation for \clarka and \leraya (from which
exact scaling laws for third order quantities follow), and review
theoretical predictions of inertial range scaling in Section
\ref{sec:MODELS}.  \add{We examine the sub-filter-scale properties of
the three regularizations} in Section \ref{sec:SFSSTUDY}.  We first
compute a fully resolved DNS of the Navier-Stokes equations at a
resolution of $1024^3$ ($\nu = 3\cdot10^{-4}$, $Re \approx 3300$, and
$R_\lambda\approx790$).  We then perform model runs with the exact
same conditions at a resolution of $384^3$.  We take $\alpha$ to be
$1/13$th the box size, which was found in an earlier study to be large
enough to exhibit both Navier-Stokes and sub-filter-scale \lansa
dynamics {\cite{PGHM+07a}}.  This large filter case is important
because it gives insight into the behavior of the models at scales
much smaller than the filter width without requiring higher resolution
than is feasible.  We compare the three regularizations as subgrid
models \add{in Section \ref{sec:SGSSTUDY}.}  Guided by {a previous
study of \lansa}\cite{PGHM+07a}, we take $\alpha$ to be $1/40$th the
box size. This choice was found to produce an optimal $\alpha-$LES (in
the sense of being optimal for the class of \lansa models, with
respect to the value of $\alpha$).  Finally, we review bounds on the
size of the attractors and use these bounds to comment on the
computational savings of the three regularizations viewed as {SGS
models}.


\section{The three regularization models}
\label{sec:MODELS}

{\subsection{\clark}}

{The incompressible Navier-Stokes equations are given in Cartesian coordinates by 
\begin{eqnarray}
\partial_t {v_i} + \partial_j({v_j}{v_i})  + \partial_i\mathcal{P}
= \add{\nu}\partial_{jj}{v_i}, \qquad
\partial_i{v_i}=0.
\label{eq:ns}
\end{eqnarray}
Filtering these equations with a convolution filter,
$L:\,{z}\rightarrow{\bar{z}}$ in which ${\bar{z}}$ (resp. ${z}$)
denotes the filtered (resp. unfiltered) field,
yields
\begin{equation}
\partial_tu_i + \partial_j(u_ju_i) + \partial_i\bar{\mathcal{P}}
+ \partial_j\bar\tau_{ij}
 = \nu \partial_{jj}u_i
 \,,
\label{eq:subgrid}
\end{equation}
in which by convention we denote $u_i\equiv\bar{v}_i$ and the Reynolds
turbulence stress tensor, $\bar\tau_{ij} = \overline{v_iv_j} -
\bar{v}_i\bar{v}_j$, represents the closure problem.
\add{Eq. (\ref{eq:subgrid}) can represent {\it either} a LES or a SGS
model.  As the difference between the two is primarily philosophical
(e.g., the scale at which filtering is applied, dissipative versus
dispersive, the factor by which computational resolution may be
decreased, etc.), we briefly define our terminology.  Many LES include
eddy-viscosity (i.e., $\partial_j\bar\tau_{ij}$ includes a $\nu_T
\partial_{jj}u_i$ term such that $\nu_T \gg \nu$).  This amounts to
approximating the $\nu =0$ problem and no finite Reynolds number can be defined.
More generally, a LES applies the filtering in the inertial range and
reduces the necessary computational linear resolution by at least an
order-of-magnitude.  Different from this previous case, a SGS model employs
a finite value of $\nu$ (and a well-defined Reynolds number) and
addresses instead the question: For a given $Re$, how far can we reduce
the computational expense while retaining as much of the detailed
large-scale properties (such as the high-order statistics) as
possible? For the case of \lans, for example, it has already been
shown that the reduction in computational expense is rather modest (a
factor of about 30).  Therefore, while calling it a SGS model is
justified by Eq. (\ref{eq:subgrid}), the label LES does not really apply.}

\add{It is the case for both LES and SGS models that though a single
filtering is indicated in Eq. (\ref{eq:subgrid}), numerical solution
implies a second filtering at the grid resolution (see, e.g.,
\cite{CWJ01}).  Systematic studies requiring a database of computed
solutions have been made in the past for LES \cite{MeSaGe2006} and for
the \lansa regularization model \cite{GH02a,GH06,PGHM+07a}.  These
studies show that the ratio of the two filter widths (i.e., the
sub-filter resolution) can affect greatly the model's performance.  To
avoid this complication, the sub-filter resolutions employed in this
study are rather large.  Determination of the optimal sub-filter
resolution is a detailed study which should be undertaken for both
\clarka and \leray, but is beyond the scope of this present paper.}

It has been shown \cite{WWV+01,CWJ01} that for all
symmetric filters possessing a finite nonzero second moment, the first term of the reconstruction series for the turbulent sub-filter stress is
\begin{equation}
\bar\tau_{ij} = -\frac{d^2\widehat{G}}{dk^2}\Big|_{k=0}\partial_k{u}_i\partial_k{u}_j +
\ldots
\label{eq:carati}
\end{equation}
where $\widehat{G}(k)$ is the Fourier transform of the convolution
kernel ($G(\vec{r})$ is the convolution kernel where
$[Lz](\vec{r})=\int G(\vec{r}-\vec{r{'}}) z(\vec{r{'}})
d^3\vec{r{'}}$).  This approximation of the subgrid stress is then
generic and is known as the Leonard tensor-diffusivity model
\cite{L74} (or, often, the Clark model \cite{CFR79}).}
{{Related to this model,} the \clarka model is \cite{CHT05}
\begin{equation}
\partial_tv_i + \mathcal H\partial_j(u_ju_i) + \partial_ip + \alpha^2
\partial_j(\partial_ku_i\partial_ku_j) = \nu \partial_{jj}v_i,
\label{eq:clark}
\end{equation}
and its subgrid stress for $\alpha\ll1$ is given by
\begin{eqnarray}
\bar\tau_{ij}^C = \mathcal H^{-1} \alpha^2
(\partial_k{u}_i\partial_k{u}_j) =
\alpha^2 (\partial_k{u}_i\partial_k{u}_j)+
\mathcal{O} (\alpha^4).
\label{eq:sgs_clark}
\end{eqnarray}
Here the filter is the inverse of a Helmholtz operator, $L = \mathcal
H^{-1} = \left( 1 - \alpha^2 \nabla^2 \right)^{-1}$.  The \clarka model 
conserves energy in the $H^1_\alpha(u)$ norm instead of the $ L^2(v)$
norm,
\begin{equation}
\frac{dE_\alpha}{dt} = -2\nu\Omega_\alpha 
\,,
\label{EQ:LANSA_BALANCE}
\end{equation}
in which the \clarka energy $E_\alpha$ is expressed as
\begin{equation}
 E_\alpha = \frac{1}{D}\int_D\frac{1}{2}(\vec{u}-\alpha^2\nabla^2\vec{u})\cdot\vec{u} d^3x
= \frac{1}{D}\int_D\frac{1}{2}\vec{v}\cdot\vec{u} d^3x
\,,
\end{equation}
and the \clarka  energy dissipation rate is given by
\begin{equation}
\Omega_\alpha = \frac{1}{D}\int_D \frac{1}{2}\boldsymbol{\omega}\cdot\bar{\boldsymbol{\omega}} d^3x
\,,
\end{equation}
where $\boldsymbol{\omega} = \vec{\nabla} \times \vec{v}$ and
$\bar{\boldsymbol{\omega}} = \vec{\nabla} \times \vec{u}$.  For $L = \mathcal
H^{-1}$, we note that $\hat{G} = (1+\alpha^2k^2)^{-1}$ which implies
that the turbulent sub-filter stress tensor for the tensor-diffusivity
model given by Eq. (\ref{eq:carati}) is
\begin{equation}
{\bar\tau^C}_{ij} = 2\alpha^2
(\partial_k{u}_i\partial_k{u}_j) + \ldots
\end{equation}
which is proportional to the \clarka stress tensor to second order in $\alpha$.
Hence, the {\sl a priori} tests of \cite{WWV+01} should apply to \clark, at least at this order.}

\subsubsection{K\'arm\'an-Howarth equation for \clark}
\label{sec:KH-clark}

{In 1938, K\'arm\'an and Howarth \cite{KH38} introduced the invariant
theory of isotropic hydrodynamic turbulence, and derived from the
Navier-Stokes equations the exact law relating the time derivative of
the two-point velocity correlation to the divergence of the
third-order correlation function. The corresponding K\'arm\'an-Howarth theorem for \lansa in the fluid case was
derived in \cite{H02c}.} The relevance of the K\'arm\'an-Howarth
theorem for the study of turbulence cannot be underestimated. As a
corollary, rigorous scaling laws in the inertial range can be
deduced. In this section, we derive these results for the \clarka
case.

For the sake of simplicity, we consider the case 
$\nu = 0$, since the dissipative terms may be added at any point 
in the derivation. 
We denote $\vec{u'}\equiv\vec{u}(\vec{x'},t)$ and begin our
investigation of the correlation dynamics by computing the ingredients
of the partial derivative $\partial_t(v_i{u'_j})$.  The \clarka  motion equation  (\ref{eq:clark})
may be rewritten as
\begin{equation}
\partial_tv_i + \partial_m(v_iu_m+u_iv_m-u_iu_m + p\delta_{im} -
\alpha^2 \partial_nu_i\partial_nu_m) = 0.
\label{eq:clarkv2}
\end{equation}
Combining Eqs. (\ref{eq:subgrid}) and (\ref{eq:sgs_clark}), we arrive at the fluctuation-velocity equation,
\begin{equation}
\partial_t{u'_j} +
\partial_m^{'}({u'_j}{u'}_m+\bar{p}'\delta_{jm}+ \alpha^2G
\otimes {\tau_{jm}^C}^{'}) = 0 \ ,
\label{eq:clarku2}
\end{equation}
where ${\bar\tau_{ij}^C} \equiv \mathcal H^{-1}\alpha^2{\tau_{ij}^C}$
($L=\mathcal H^{-1}$).
Multiplying Eq. (\ref{eq:clarkv2}) by ${u'_j}$ and Eq. (\ref{eq:clarku2}) by $v_i$, then adding the result yields
\begin{eqnarray}
\partial_t\langle v_i{u'_j}\rangle = \frac{\partial}{\partial{r_m}}
\langle (v_iu_m+u_iv_m-u_iu_m - \alpha^2 \partial_nu_i\partial_nu_m) {u'_j}
\rangle + \frac{\partial}{\partial{r_m}}
\langle p{u'_j}\delta_{im} - \bar{p}'v_i\delta_{jm} \rangle \nonumber \\
- \frac{\partial}{\partial{r_m}}
\langle ({u'_j}{u'}_m + \alpha^2G\otimes {{\tau_{jm}^C}'}) v_i\rangle
\,,
\label{fluct-eqn}
\end{eqnarray}
where we have used statistical homogeneity
\begin{equation}
\frac{\partial}{\partial r_m} \left< \cdot \right> = 
    \frac{\partial}{\partial x'_m} \left< \cdot \right> = 
    - \frac{\partial}{\partial x_m} \left< \cdot \right> 
    \,.
\end{equation}
We symmetrize Eq. (\ref{fluct-eqn}) in the indices $i,j$ by adding 
the corresponding equation for $\partial_t \langle v_j u'_i \rangle$. 
We then use homogeneity again as 
\begin{equation}
\left< v_i u'_j u'_m + v_j u'_i u'_m \right> = 
    - \left< v'_i {u}_j {u}_m + v'_j {u}_i {u}_m \right>,
\end{equation}
and define the tensors
\begin{eqnarray}
{{\cal Q}}^C_{ij} &=& \left< v_i u'_j + v_j u'_i 
    \right> , \\
{{\cal T}}^C_{ijm} &=& < \left(v_i u'_j + 
    v_j u'_i + {v'_i} {u}_j +
    {v'_j} {u}_i - u_iu'_j - u_ju'_i \right) {u}_{m} + \nonumber \\
 \left( u_iu'_j+u_ju'_i\right)v_m >
 , \label{eqT} \\
\Pi^C_{ijm} &=& \left< \left( p u'_j - \bar{p}' 
    {v}_j \right) \delta_{im} + \left( p u'_i - 
    \bar{p}' {v}_i \right) \delta_{jm} \right> , \\
{{\cal S}}^C_{ijm} &=& < \left( \partial_nu_i\partial_nu_m\right) u'_j
 + \left( \partial_nu_j\partial_nu_m\right) u'_i
    + G \otimes {{\tau_{jm}^C}'} v_i + \nonumber \\
G \otimes {{\tau_{im}^C}'} 
    v_j > . \label{eqS}
\end{eqnarray}
We can drop $\Pi^C_{ijm}$ because the terms with the pressures $p$
and $\bar{p}'$ vanish everywhere, as follows from the arguments of
isotropy \cite{KH38}. Finally, we obtain
\begin{equation}
\partial_t {{\cal Q}}^C_{ij} = \frac{\partial}{\partial r_m} \left( 
    {{\cal T}}^C_{ijm} - \alpha^2 {{\cal S}}^C_{ijm} \right) .
\label{eq:KH1}
\end{equation}
This is the K\'arm\'an-Howarth equation for \clarka (compare to {Eq. (3.8) in \cite{H02c}} for \lans).

By dimensional analysis, {the energy dissipation rate in \clarka  
is $\epsilon_\alpha^C \sim \partial_t {{\cal Q}^C}_{ij}$ and} 
Eq. (\ref{eq:KH1}) implies
\begin{equation}
\varepsilon_\alpha^C \sim \frac{1}{l}(vu^2 + u^3 + \frac{\alpha^2}{l^2}u^3).
\label{EQ:FULL_CLARK}
\end{equation}
For large scales ($l \gg \alpha$), we recover the Navier-Stokes
scaling known as the four-fifths law, $<(\delta v_{\|}(l))^3> \sim
\varepsilon l$ {\cite{F95}}.  Here, $\delta v_{\|}(l) \equiv
[\vec{v}(\vec{x+l})-\vec{v}(\vec{x})]\cdot\vec{l}/l$ is the
longitudinal increment of $\vec{v}$. {Strictly speaking,} the 
four-fifths law expresses
that the third-order longitudinal structure function of $\vec{v}$,
${S}_3^v (l) \equiv \langle(\delta v_{\|})^3\rangle$, is given in
the inertial range in terms of the mean energy dissipation per unit
mass $\varepsilon$ by
\begin{equation}
{S}_3^v = -\frac{4}{5}\varepsilon l,
\label{EQ:FOURFIFTHS}
\end{equation}
or, equivalently, that the flux of energy across scales in the
 inertial range is constant.  We also recover the {Kolmogorov 1941
 \cite{K41a,K41b,K41c} (hereafter, K41)} energy spectrum, $E(k)k \sim
 v^2 \sim \varepsilon^{2/3}l^{2/3}$ or, equivalently,
\begin{equation}
E(k) \sim \varepsilon^{2/3}k^{-5/3}.
\label{eq:K41}
\end{equation}
For sub-filter scales ($l \ll \alpha$), we have $u\sim vl^2/\alpha^2$
and the first and third right-hand terms in Eq. (\ref{EQ:FULL_CLARK})
are equivalent.  In this case, we are left with two different possible
scalings depending on the prefactors in Eq. (\ref{EQ:FULL_CLARK}).  If
the first (or third) right-hand term is dominant, our scaling law
becomes
\begin{equation}
\left<(\delta u_{\|}(l))^2(\delta v_{\|}(l))\right> \sim
\varepsilon_\alpha^C l.
\label{EQ:LCUBE}
\end{equation}
For our sub-filter scale energy spectrum we would then have
$E_\alpha^C(k)k \sim uv \sim (\varepsilon_\alpha^C)^{2/3}\alpha^{2/3}$,
or, equivalently,
\begin{equation}
E_\alpha^C(k)\sim (\varepsilon_\alpha^C)^{2/3}\alpha^{2/3}k^{-1}.
\label{EQ:CLARKA_SPECTRUM}
\end{equation}
This result is the same as for the $\alpha-$model \cite{FHT01}.  If,
however, the second right-hand term in Eq. (\ref{EQ:FULL_CLARK}) is
dominant, then the K41 results are recovered, with $\vec{u}$ substituted for $\vec{v}$.  In that case, one finds the alternative \clarka sub-filter scale spectral energy scaling, 
\begin{equation}
E_\alpha^C(k)\sim k^{1/3}
\,.
\label{EQ:CLARKA_SPECTRUM2}
\end{equation}

\subsubsection{Phenomenological arguments for \clarka inertial range scaling}
\label{sec:pheno_clark}

{We review here the derivation by dimensional analysis of the spectrum which follows the scaling ideas originally due to Kraichnan \cite{K67} and
which is developed more fully in Ref. \cite{CHT05}.  In examining the
nonlinear terms in Eq. (\ref{eq:clark}), it is not entirely clear
which of three possible scales for the average velocity for an eddy of
size $k^{-1}$,
\begin{equation}
U_k^{(0)} = \left( \frac{1}{D} \int_D |\vec{v}_k|^2 d^3x \right)^{1/2},
\label{eq:Uk0}
\end{equation}
\begin{equation}
U_k^{(1)} = \left( \frac{1}{D} \int_D \vec{u}_k\cdot\vec{v}_k d^3x \right)^{1/2},
\label{eq:Uk1}
\end{equation}
or
\begin{equation}
U_k^{(2)} = \left( \frac{1}{D} \int_D |\vec{u}_k|^2 d^3x \right)^{1/2}
\label{eq:Uk2}
\end{equation}
should result.  Therefore, three corresponding ``turnover times'', $\mathfrak{t}_k$,
for such an eddy may be proposed
\begin{equation}
\mathfrak{t}_k^{(n)} \sim 1/(kU_k^{(n)})
\quad\hbox{with}\quad
n=0,1,2
\,.
\label{eq:TURNOVER}
\end{equation}
The term ``turnover time'' is used advisedly here, since only the velocity $U_k^{(2)}$ is composed of the fluid transport velocity.
We define the (omnidirectional) spectral energy density, $E_\alpha(k)$, from the relation
\begin{equation}
E_\alpha = \int_0^\infty \oint E_\alpha(\vec{k}) d\sigma d\vec{k} =
\int_0^\infty E_\alpha(k) dk
\,.
\end{equation}
Since, $\vec{u}_k\cdot\vec{u}_k =
\vec{u}_k\cdot\vec{v}_k/({1+\alpha^2k^2}) =
{E_\alpha(k)}/({1+\alpha^2k^2})$, we have
\begin{equation}
U_k^{(n)} \sim \left( \int {E_\alpha(k)}({1+\alpha^2k^2})^{(1-n)} dk \right)^{1/2}
\sim \left(  {kE_\alpha(k)}({1+\alpha^2k^2})^{(1-n)} \right)^{1/2}_.
\end{equation}
Then, the total energy dissipation rate, $\varepsilon_\alpha^C$ is related to the
spectral energy density by
\begin{equation}
\varepsilon_\alpha \sim (\mathfrak{t}_k^{(n)})^{-1} \int E_\alpha(k) dk
\sim k^2 U_k^{(n)} E_\alpha(k) \sim {k^{5/2}E_\alpha^C(k)^{3/2}}{(1+\alpha^2k^2)^{(1-n)/2}},
\end{equation}
which yields, finally, the predicted energy spectra for \clark,
$E_\alpha^C(k)$,
\begin{equation}
E_\alpha^C(k)
\sim (\varepsilon_\alpha^C)^{2/3} k^{-5/3}(1+\alpha^2k^2)^{(n-1)/3}.
\label{EQ:CLARKA_SPECTRUMn}
\end{equation}
For scales much larger than $\alpha$ ($\alpha k \ll 1$) the Kolmogorov scaling for Navier-stokes is recovered,
\begin{equation}
E_\alpha^C(k)
\sim  (\varepsilon_\alpha^C)^{2/3}k^{-5/3},
\end{equation}
whereas for scales much smaller ($\alpha k \gg 1$), the spectrum becomes
\begin{equation}
E_\alpha^C(k)
\sim  (\varepsilon_\alpha^C)^{2/3}\alpha^{(2(n-1)/3}k^{(2n-7)/3}.
\end{equation}
These arguments}
constrain the \clarka sub-filter scale spectrum to lie
between $k^{-1}$ and $k^{-7/3}$.

{\subsection{\leray}}

{The Leray model in Cartesian coordinates is
\begin{eqnarray}
\partial_tv_i + \partial_j(u_jv_i) + \partial_iP = \nu \partial_{jj}v_i
\qquad \partial_iv_i=0,
\label{eq:leray}
\end{eqnarray}
where the flow is advected by a smoothed velocity, $\vec{u}$.  By
comparison with Eq. (\ref{eq:subgrid}) we see that the Leray model
approximates the subgrid stress as $\bar\tau_{ij}^L = L(u_jv_i) -
u_ju_i$, or, with $L = \mathcal H^{-1}$,
\begin{equation}
{\bar\tau_{ij}^L} = \mathcal H^{-1}\alpha^2 (\partial_k{u}_i\partial_k{u}_j +
\partial_k{u}_i\partial_j{u}_k).
\label{eq:sgs_leray}
\end{equation}
As has been noted previously \cite{HN03}, the subgrid stress of
\clark\, in Eq.  (\ref{eq:sgs_clark}) is a truncation of the subgrid
stress of \leray, in Eq. (\ref{eq:sgs_leray}).  For \leray, the $L^2(v)$
norm is the quadratic invariant that is identified with energy,
\begin{equation}
\frac{dE}{dt} =-2\nu\Omega,
\label{EQ:LERAY_BALANCE}
\end{equation}
where
\begin{equation}
 E = \frac{1}{D}\int_D\frac{1}{2}|\vec{v}|^2 d^3x,
\end{equation}
and
\begin{equation}
\Omega = \frac{1}{D}\int_D \frac{1}{2}|\boldsymbol{\omega}|^2 d^3x.
\end{equation}
As was
pointed out in \cite{RJH06}, the incompressibility of the velocity
field $\vec{v}$ only implies a divergenceless filtered velocity
$\vec{u}$ under certain boundary conditions for \leray.  When
$\partial_iu_i \ne 0$, the energy $E = \frac{1}{2}\int_D |\vec{v}|^2$ is no
longer conserved (helicity and Kelvin's theorem are not conserved for
\leray).  In our numerical study, we employ periodic boundary
conditions, for which $\partial_iv_i = 0$ implies $\partial_iu_i = 0$
and \leraya conserves energy in the usual sense of  $ L^2(v)$.}

\subsubsection{K\'arm\'an-Howarth equation for \leray}
\label{sec:KH-leray}

In this section we derive the K\'arm\'an-Howarth equation for the
\leraya case.  Following
Section \ref{sec:KH-clark}, we begin our
investigation of the correlation dynamics by computing the ingredients
of the partial derivative $\partial_t(v_i{v'_j})$.  Eq. (\ref{eq:leray})
may be rewritten as:
\begin{equation}
\partial_tv_i + \partial_m(v_iu_m + P\delta_{im}) = 0.
\label{eq:lerayv2}
\end{equation}
Multiplying Eq. (\ref{eq:lerayv2}) by ${v'_j}$ yields
\begin{eqnarray}
\partial_t\langle v_i{v'_j}\rangle = \frac{\partial}{\partial{r_m}}
\langle v_iu_m{v'_j}
\rangle + \frac{\partial}{\partial{r_m}}
\langle P{v'_j}\delta_{im} \rangle.
\end{eqnarray}
We can make this equation symmetric in the indices $i,j$, by adding  the equation for $\partial_t \langle v_j v'_i \rangle$. 
We define the tensors
\begin{eqnarray}
{{\cal Q}}^L_{ij} &=& \left< v_i v'_j + v_j v'_i \right> , \\ {{\cal
    T}}^L_{ijm} &=& \left< \left(v_i v'_j + v_j v'_i \right) {u}_{m}
    \right> , \label{eqT2} \\ \Pi^L_{ijm} &=& \left< P v'_j
    \delta_{im} + P v'_i \delta_{jm} \right>. \label{eqSleray}
\end{eqnarray}
Again, we may drop $\Pi^L_{ijm}$ because the terms with the
pressure $P$ vanish everywhere and thereby obtain
\begin{equation}
\partial_t {{\cal Q}}^L_{ij} = \frac{\partial}{\partial r_m} 
    {{\cal T}}^L_{ijm}.
\label{eq:KH1leray}
\end{equation}
This is the K\'arm\'an-Howarth equation for \leray.

The energy dissipation rate for \leray\, is denoted by $\varepsilon^L$ and it satisfies
$\varepsilon^L \sim \partial_t{{\cal Q}^L}_{ij}$.  By dimensional
analysis, Eq. (\ref{eq:KH1leray}) implies
\begin{equation}
\varepsilon^L \sim \frac{1}{l}v^2u.
\label{EQ:FULL_SCALEleray}
\end{equation}
For large scales ($l \gg \alpha$), we recover the Navier-Stokes
scaling, Eqs. (\ref{EQ:FOURFIFTHS}) and (\ref{eq:K41}).  For sub-filter
scales ($l \ll \alpha$) our scaling law becomes
\begin{equation}
\left<(\delta v_{\|}(l))^2(\delta u_{\|}(l))\right> \sim
\varepsilon^L l.
\label{EQ:LCUBEleray}
\end{equation}
For our small scale energy spectrum we would then have
$E^L(k)k \sim v^2 \sim (\varepsilon^L)^{2/3}\alpha^{4/3}
k^{2/3}$ (where we employed $u\sim vl^2/\alpha^2$),
or, equivalently, cf. Eq. (\ref{EQ:CLARKA_SPECTRUM2}),
\begin{equation}
E^L(k)\sim (\varepsilon^L)^{2/3}\alpha^{4/3}k^{-1/3}.
\label{EQ:LERAY_SPECTRUM}
\end{equation}

{\subsubsection{Phenomenological arguments for \leraya inertial range scaling}}

{We review here the derivation by dimensional analysis of the spectrum for
\leraya as we did for \clarka in Section \ref{sec:pheno_clark}.  This
analysis is developed more fully in Ref. \cite{CHO+05}.  We argue
again that there are three possible scales for the average velocity
for an eddy of size $k^{-1}$, Eqs. (\ref{eq:Uk0}), (\ref{eq:Uk1}), and
(\ref{eq:Uk2}), with the turn-over time, $\mathfrak{t}_k^{(n)}$ given by
Eq. (\ref{eq:TURNOVER}). Since, ${u}_k^2 = {v}_k^2/({1+\alpha^2k^2})^2
= {E(k)}/({1+\alpha^2k^2})^2$, we have
\begin{equation}
U_k^{(n)} \sim \left( \int {E(k)}({1+\alpha^2k^2})^{-n} dk \right)^{1/2}
\sim \left(  {kE(k)}({1+\alpha^2k^2})^{-n} \right)^{1/2}_.
\end{equation}
Then, the total energy dissipation rate, $\varepsilon^L$ is related to the
spectral energy density by
\begin{equation}
\varepsilon^L \sim (\mathfrak{t}_k^{(n)})^{-1} \int E(k) dk
\sim k^2 U_k^{(n)} E(k) \sim {k^{5/2}E(k)^{3/2}}{(1+\alpha^2k^2)^{-n/2}},
\end{equation}
which yields, finally, the predicted energy spectra for \leray,
$E^L(k)$,
\begin{equation}
E^L(k)
\sim (\varepsilon^L)^{2/3} k^{-5/3}(1+\alpha^2k^2)^{n/3}.
\label{EQ:LERAY_SPECTRUM2}
\end{equation}
For scales much larger than $\alpha$ ($\alpha k \ll 1$) the K41
spectrum is recovered, Eq. (\ref{eq:K41}), and for scales much smaller
($\alpha k \gg 1$) the spectrum is
\begin{equation}
E^L(k)
\sim  (\varepsilon^L)^{2/3}\alpha^{2n/3}k^{(2n-5)/3}.
\end{equation}
These arguments constrain the \leraya sub-filter scale spectrum to lie
between $k^{-1/3}$ and $k^{-5/3}$.}

\subsection{\lans}

{\lansa is given by
\begin{eqnarray} \partial_tv_i +
\partial_j(u_jv_i) + \partial_i\pi + v_j\partial_iu_j = \nu
\partial_{jj}v_i, \qquad \partial_iv_i = 0,
\label{eq:lans}
\end{eqnarray}
For \lans, the usual choice of filter is again $L = \mathcal H.^{-1}$
With this filter, the subgrid stress tensor is given by
\begin{eqnarray}
{\bar\tau_{ij}^\alpha} = \mathcal H^{-1} \alpha^2 (\partial_m{u}_i\partial_m{u}_j +
\partial_m{u}_i\partial_j{u}_m - \partial_i{u}_m\partial_j{u}_m). 
\label{eq:sgs_lans}
\end{eqnarray}
As has been previously noted \cite{HN03}, the subgrid stress of
\leray, Eq. (\ref{eq:sgs_leray}), is a truncation of the subgrid
stress of \lansa Eq. (\ref{eq:sgs_lans}).  Like \clark, energy is
conserved in the $ H^1_\alpha(u)$ norm instead of the $ L^2(v)$ norm.
Additionally, \lansa is the only model of the three examined here that
conserves \add{a
form of} the helicity (and Kelvin's circulation theorem).}

{For \lansa in the fluid case the K\'arm\'an-Howarth theorem was
derived in \cite{H02c}.  We summarize here the dimensional analysis
argument for the \lansa inertial range scaling that follows from this
theorem, beginning from Equation (3.8) in Ref. \cite{H02c}.  In the
statistically isotropic and homogeneous case, without external forces
and with $\nu=0$, taking the dot product of Eq. (\ref{eq:lans}) with
$u{'}_j$ yields the equation}

\begin{equation}
\partial_t {\mathcal{Q}}^\alpha_{ij} = \frac{\partial}{\partial{r_m}}\left(
{\mathcal T}^\alpha_{ijm} - \alpha^2 {\mathcal S}^\alpha_{ijm} \right).
\label{eq:Holm38}
\end{equation}
The trace of this equation is the Fourier transform of the detailed
energy balance for \lans;
\begin{equation}
 {\mathcal Q}^\alpha_{ij} = \left< v_iu_j^{'} + v_ju_i^{'}\right> 
\end{equation}
is the second-order correlation tensor while
\begin{equation}
 {\mathcal T}^\alpha_{ijm} = \left< (v_iu_j^{'} +v_ju_i^{'} +v_i^{'}u_j +v_j^{'}u_i )u_m\right>,
\end{equation}
and
\begin{equation}
{\mathcal S}^\alpha_{ijm} = \left< (\partial_mu_l\partial_iu_l)u_j^{'}
+(\partial_mu_l\partial_ju_l)u_i^{'}
+(G \otimes {{{\tau_{jm}^\alpha}'}} )v_i\\
+(G \otimes {{{\tau_{im}^\alpha}'}} )v_j\right>,
\end{equation}
are the third-order correlation tensors for \lansa and
${\bar\tau}^\alpha_{ij} = \mathcal H^{-1}\alpha^2 \tau^\alpha_{ij}$
is the sub-filter scale stress tensor.
{For $\alpha=0$ this reduces to the well-known relation
derived by K\'arm\'an and Howarth.}  The energy dissipation rate for
\lans, $\varepsilon_\alpha$, satisfies $\varepsilon_\alpha \propto
\partial_t {\mathcal Q^\alpha}_{ij}$.  By dimensional analysis in
Eq. (\ref{eq:Holm38}) we arrive at
\begin{equation}
\varepsilon_\alpha \sim \frac{1}{l}(vu^2 + \frac{\alpha^2}{l^2}u^3).
\label{EQ:FULL_SCALE}
\end{equation}
For large scales ($l \gg \alpha$), we recover the Navier-Stokes
scaling Eqs. (\ref{EQ:FOURFIFTHS}) and (\ref{eq:K41}).  For sub-filter
scales ($l \ll \alpha$) our scaling law becomes Eq. (\ref{EQ:LCUBE})
and our sub-filter scale spectra is given by
\begin{equation}
E_\alpha(k)\sim \varepsilon_\alpha^{2/3}\alpha^{2/3}k^{-1}.
\label{EQ:LANSA_SPECTRUM}
\end{equation}
In this case, by the phenomenological arguments, we know that eddies
of size $k^{-1}$ are advected by the smoothed velocity,
Eq. (\ref{eq:Uk2}).  This scaling is confirmed in {Ref. \cite{PGHM+07a}}
but it coexists with a $k^1$ energy spectrum corresponding to ``enslaved rigid
bodies'' or ``polymerized'' portions of fluid which do not
contribute to the turbulent energy cascade.

\section{Sub-filter-scale physics}
\label{sec:SFSSTUDY}

\add{Only by examining the sub-filter scales can we hope to derive
new, improved models, and, ultimately, to gain an understanding of
turbulence.  A knowledge of the differences between closures and
Navier-Stokes is fundamental to enable the derivation of better
physical models of turbulence at small scales.  In this section, then,
we will be interested in both the similarities and the differences
between the regularizations and Navier-Stokes.  A more immediate goal
of predicting the computational savings at higher Reynolds numbers can
be achieved through the correct prediction of the scaling at small
scales.}

To this end, we compute numerical solutions to Eqs. (\ref{eq:ns}),
(\ref{eq:clark}), (\ref{eq:leray}), and (\ref{eq:lans}) in a
three-dimensional (3D) cube with periodic boundary conditions using a
parallel pseudospectral code \cite{GMD05,GMD05b}.  We employ a
Taylor-Green forcing \cite{TG37}{,
\begin{equation}
F = \left[ \begin{array}{c}
    \sin k_0x \cos k_0y \cos k_0z \\
  - \cos k_0x \sin k_0y \cos k_0z \\
  0
  \end{array}\right]
\label{EQ:TG}
\end{equation}
(with $k_0=2$), and employ dynamic control \cite{MPM+05} to
maintain a nearly constant energy with time.  The Taylor-Green
forcing, Eq. (\ref{EQ:TG}), is not a solution of the Euler's
equations, and as a result small scales are generated rapidly.  The
resulting flow models the fluid between counter-rotating cylinders
\cite{B90} and it has been widely used to study turbulence, including
studies in the context of the generation of magnetic fields through
dynamo instability \cite{PMM+05}.  We define the Taylor microscale as
$ \lambda = 2\pi\sqrt{\langle v^2\rangle/\langle \omega^2\rangle}, $
and the mean velocity fluctuation as $ v_{rms} = \left( 2\int_0^\infty
E(k) dk\right)^{1/2}$. The Taylor microscale Reynolds number is
defined by $ R_\lambda = {v_{rms}\lambda}/{\nu} $ and the Reynolds
number based on a unit length is $ Re = v_{rms}/{\nu}$.}

\add{The \clark, \leray, and \lansa equations (as well as other SGS
  models based on spectral filters) are easy to implement in spectral
  or pseudospectral methods.  As an example, in Fourier based
  pseudospectral methods, the Helmholtz differential operator can be
  inverted to obtain $\widehat{\mathcal H}^{-1}(k) = (1+\alpha^2
  k^2)^{-1}$, where the hat denotes Fourier transformed. In this way,
  the filter reduces to an algebraic operation and
  Eqs. (\ref{eq:clark}), (\ref{eq:leray}), and (\ref{eq:lans}) can be
  solved numerically at almost no extra cost. If other numerical
  methods are used, the inversion can be circumvented for example by
  expanding the inverse of the Helmholtz operator into higher orders of
  the Laplacian operator \cite{ZM04,S06}.}

To compare the three regularizations (\clark, \leray, and \lans) we
compute a fully resolved DNS of the Navier-Stokes equations at a
resolution of $1024^3$ ($\nu = 3\cdot10^{-4}$, $Re\approx3300$) and
model runs with the exact same conditions at a resolution of $384^3$.
The details of the flow dynamics of the DNS have already been given
\cite{AMP05b,MAP06}.  In particular, the Reynolds number based on the
integral scale $\mathfrak L \equiv 2\pi\int E(k)k^{-1}dk/E\approx1.2$
(where $E$ is the total energy) is $Re_{\mathfrak L}=U\mathfrak
L/\nu\approx3900$, where $U$ is the {\sl r.m.s.} velocity and the
Reynolds number based on the Taylor scale is $R_\lambda\approx790$.
The DNS was run for nine turnover times $(\mathfrak L/U)$ (in the
following results, time $t$ is in units of the turnover time). We
employ a filter width of
$\alpha = 2\pi / 13$ for which \lansa exhibits both Navier-Stokes and
\lansa inertial ranges in the third-order structure function
{\cite{PGHM+07a}}.  
From these we hope to obtain the behavior of
the models for scales much smaller than $\alpha$ 

\begin{figure}[htbp] 
    \includegraphics[width=8.95cm]{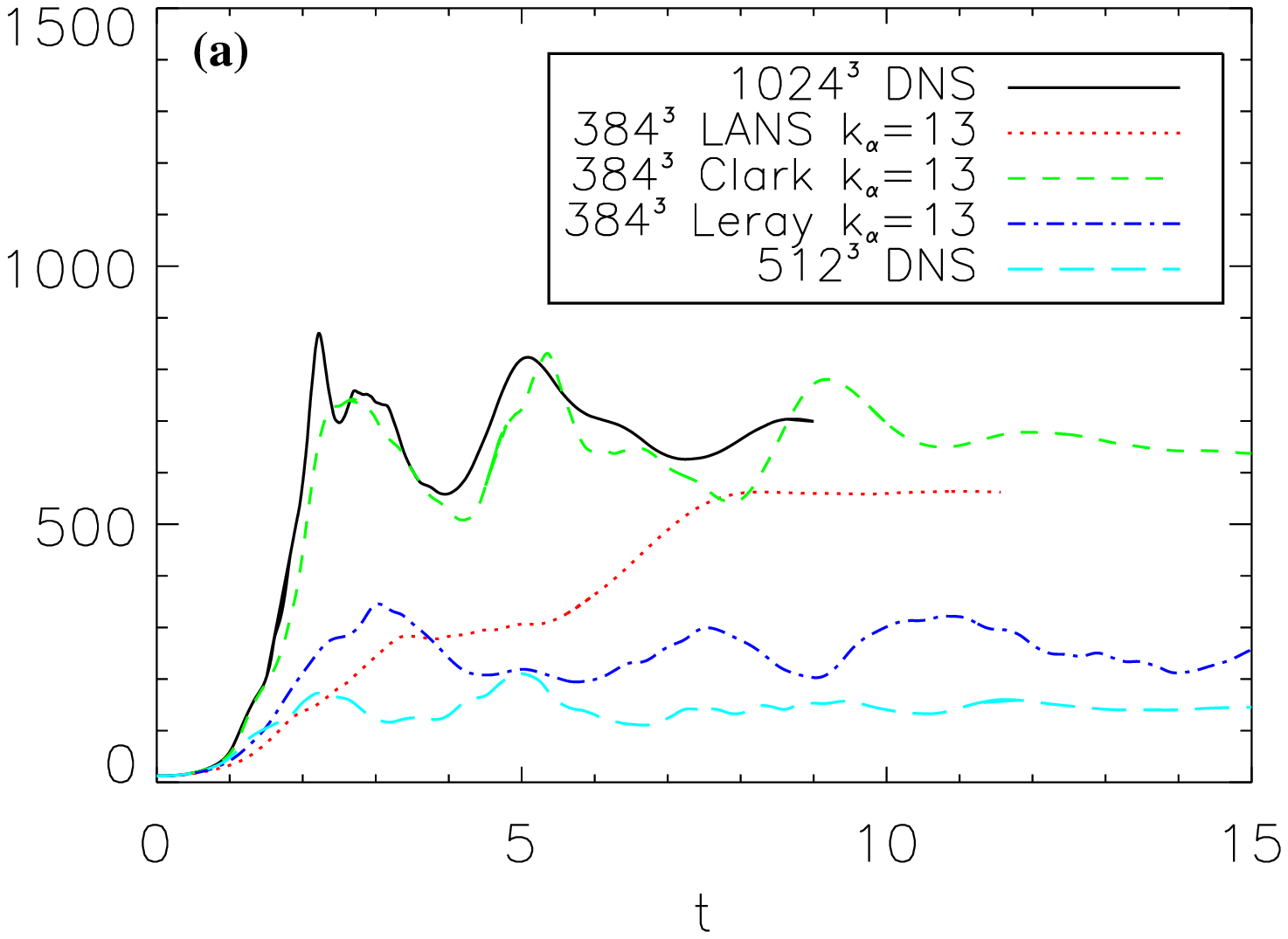} 
  \caption[Time evolution of dissipation] {(Color online) Time evolution of 
  {the enstrophy ($\left<\omega^2\right>$ for \leraya and in the DNS, 
  $\left<\boldsymbol{\omega}\cdot\boldsymbol{\overline{\omega}}\right>$ for 
  \lansa and \clark)}. DNS ($Re\approx3300$) is shown as solid black 
  lines, \lansa as dotted
  red, \clarka as dashed green, and \leraya as blue dash-dotted.
  The cyan long-dashed line represents a $512^3$ DNS ($Re
  \approx 1300$, $R_\lambda \approx 490$). Here each run is calculated
  only until it reaches a statistical steady state.  \leraya reduces
  the dissipation, $\varepsilon = \nu\left< \omega^2 \right>$, and
  increases the time scale to reach a statistical turbulent
  steady-state.  Both effects are greater as $\alpha$ is increased. By
  comparison with the $Re \approx 1300$ run, we see that these two
  effects are consistent with a reduced effective Reynolds number.  A
  smaller reduction in flux (but not an increase in time to
  steady-state) is also observed for \lansa and is likely related to
  its rigid bodies.}
  \label{fig:ENSTROPHY}
\thesis{\end{center}}
\end{figure}

In Fig. \ref{fig:ENSTROPHY} we present the time evolution of the
enstrophy {($\left<\omega^2\right>$ for \leraya and DNS,}
$\left<\boldsymbol{\omega}\cdot\boldsymbol{\overline{\omega}}\right>$
for \lansa and \clark), which is proportional to the dissipation
{($\varepsilon = \nu \left< \omega^2\right>$ or $\varepsilon_\alpha =
\nu
\left<\boldsymbol{\omega}\cdot\boldsymbol{\overline{\omega}}\right>$
depending on the case).}  Also shown is a well-resolved $512^3$ DNS of
a less turbulent flow ($\nu=1.5\cdot10^{-3}$, $Re\approx1300$,
$R_\lambda \approx 490$, (cyan online) long-dashed line).  Here each
run is calculated only until it reaches a statistically steady state.
We see that the dissipation is \add{greatly} reduced and the
time-scale \add{to reach a statistically steady state} is increased
for \leray.  We see by comparison with the $Re\approx1300$ DNS that
\add{this reduced dissipation could result from} a reduced effective
Reynolds number in the \leraya run.
For \lans, the dissipation is decreased
although the time to reach steady state is not increased. This is
probably related to the enslavement of its rigid body regions which
would have no internal dissipation.  Of the three models, \clarka
\add{most resembles} the total dissipation for a large range of
$\alpha$.  Indeed, as it is the order $\alpha^2$ approximation of
Navier-Stokes, this dissipation behavior for \clarka\, may continue to
hold until $\alpha$ becomes quite large.

\rem{
\begin{figure}[htbp]\thesis{\begin{center}\leavevmode}
    \includegraphics[width=8.95\thesis{11}cm]{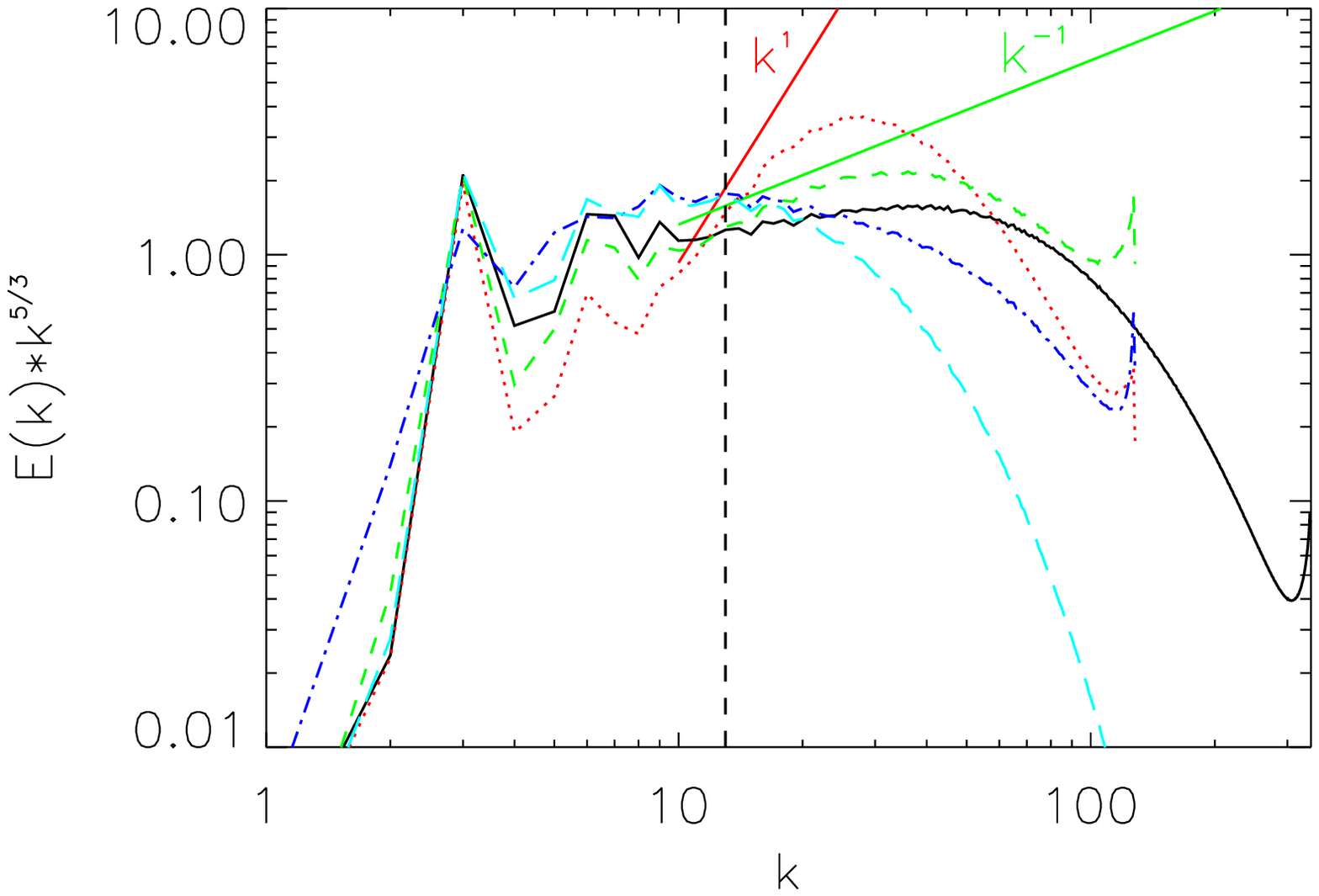} 
  \caption[Averaged compensated spectra] {(Color online) Spectra compensated by K41 for $1024^3$ DNS ($Re\approx3300$) averaged
over $t=[8.25,9]$.  Labels are as in Fig. \ref{fig:ENSTROPHY}.
The vertical dashed line indicates $k_\alpha \equiv 2\pi/\alpha$.  Compensated spectra
for $384^3$ \lansa averaged over $t\in[10.8,11.6]$, for $384^3$
\clarka over $t\in[11.8,12.6]$, for $384^3$ \leraya over
$t\in[26.3,27]$, and for the $Re\approx1300$ DNS over
$t\in[18.1,18.9]$.  Due to the large disparity in times to reach a
turbulent steady-state, the time intervals chosen to average over also
differ greatly.  \clarka \add{approximates} the predicted $k^{-1}$ spectrum,
Eq. (\ref{EQ:CLARKA_SPECTRUM}), and not $k^{1/3}$,
Eq. (\ref{EQ:CLARKA_SPECTRUM2}), nor another possible spectrum,
Eq. (\ref{EQ:CLARKA_SPECTRUMn}).  The spectrum of \leraya is very
similar (for $k \in [5,20]$) to that of the $Re\approx1300$ DNS.  The \sunu{positive but shallower than $k^1$} \lansa
spectrum observed here has previously been reported \sunu{(see text)}.  \add{The spectra of all
three regularizations clearly differ from that of Navier-Stokes.}}
  \label{fig:EAVG}
\thesis{\end{center}}
\end{figure}
}

Figure \ref{fig:EAVG} shows a comparison of the energy spectrum 
at the turbulent steady state, for all the runs in Fig. \ref{fig:ENSTROPHY}. 
The isotropic energy spectra are calculated as follows,
\begin{equation}
E(k) = \sum_{k_\textrm{eff} \ge k-\frac{1}{2}}^{k_\textrm{eff} < k+\frac{1}{2}}
v_x^2(k_\textrm{eff})+v_y^2(k_\textrm{eff})+v_z^2(k_\textrm{eff})
\end{equation}
where $k_\textrm{eff} = \sqrt{k_x^2+k_y^2+k_z^2}$ (the $
H_1^\alpha(u)$ norm is employed for \clarka and \lans).  The length
scale $\alpha$ is indicated by a vertical dashed line and the plotted
energy spectra are compensated by $k^{5/3}$ (i.e., leading to a flat
K41 $k^{-5/3}$ spectrum).  \add{The energy flux in the DNS is constant
  in a wide range of scales, but the compensated spectrum has a more
  complex structure. The salient features of this spectrum are
  well-known from previous studies \cite{KIY+03}. Small scales before the
  dissipative range show the so-called bottleneck effect with a slope
  shallower than $k^{-5/3}$. On the other hand, larger scales have a
  tendency to develop a spectrum slightly steeper than $k^{-5/3}$
  because of intermittency corrections, an effect that becomes clear
  in the simulation performed at larger spatial resolution on a grid
  of $4096^3$ points \cite{KIY+03}.}  From Fig. \ref{fig:EAVG} it is
clear that the \clarka spectral behavior is close to the predicted
$k^{-1}$ spectrum, rather than the $k^{1/3}$ from
Eq. (\ref{EQ:CLARKA_SPECTRUM2}), or the other possible spectrum from
Eq. (\ref{EQ:CLARKA_SPECTRUMn}).  Likewise, the \sunu{positive $k^{0.2}$ \lansa
spectrum observed here approaches $k^1$ with increasing resolution as the sub-filter scales are fully resolved} \cite{PGHM+07a}.  \leray, on the
other hand, possesses a very steep sub-filter scale spectrum as well
as enhanced large-scale energy as has been previously observed
\cite{GH06}.  \add{The results indicate that solutions to \leraya are
  the most strongly regularized of the three regularizations.}

The spectrum of \leraya in Fig. \ref{fig:EAVG} gives  
a good approximation to the $Re \approx 1300$ DNS in the range $k\in[5,20]$
(i.e., to $\nu=1.5\cdot10^{-3}$ rather than to $\nu=3\cdot10^{-4}$ which was employed). This result, \leray's
increased characteristic time scales and its reduced dissipation imply 
that the \leraya model is operating at a much lower effective Reynolds 
number. This is also clear from the rapid drop in the spectrum at 
small scales, shown in Fig. \ref{fig:EAVG}. Indeed, we can build an 
effective Reynolds number in the large scales as 
$R_\textrm{eff} = \epsilon^{1/3} \mathfrak{L}^{4/3} / \nu$. Since 
$\mathfrak{L}$ is controlled in this simulation by the forcing scale, 
the drop in the dissipation rate implies a reduced nonlinearity in \leray.
This is also consistent with a direct comparison of the nonlinear 
terms in \leraya with, for instance, \lans.  The nonlinear terms 
in \lans, Eq. (\ref{eq:lans}), may be written as 
${\bf u} \cdot \nabla {\bf v}+ \nabla {\bf u}^T \cdot {\bf v}$ 
(where the suffix $T$ denotes a transposition), while
the nonlinear term in \leray, Eq. (\ref{eq:leray}), is only
$\vec{u}\cdot\vec{\nabla}\vec{v}$.
{Both nonlinear terms in \lansa are of order $O(1)$; so the absence of one of the nonlinear terms in \leraya \add{could} be understood as a reduction in the nonlinearity.}

\rem{
\begin{figure}[htbp]\thesis{\begin{center}\leavevmode}
    \includegraphics[width=8.95\thesis{12.5}cm]{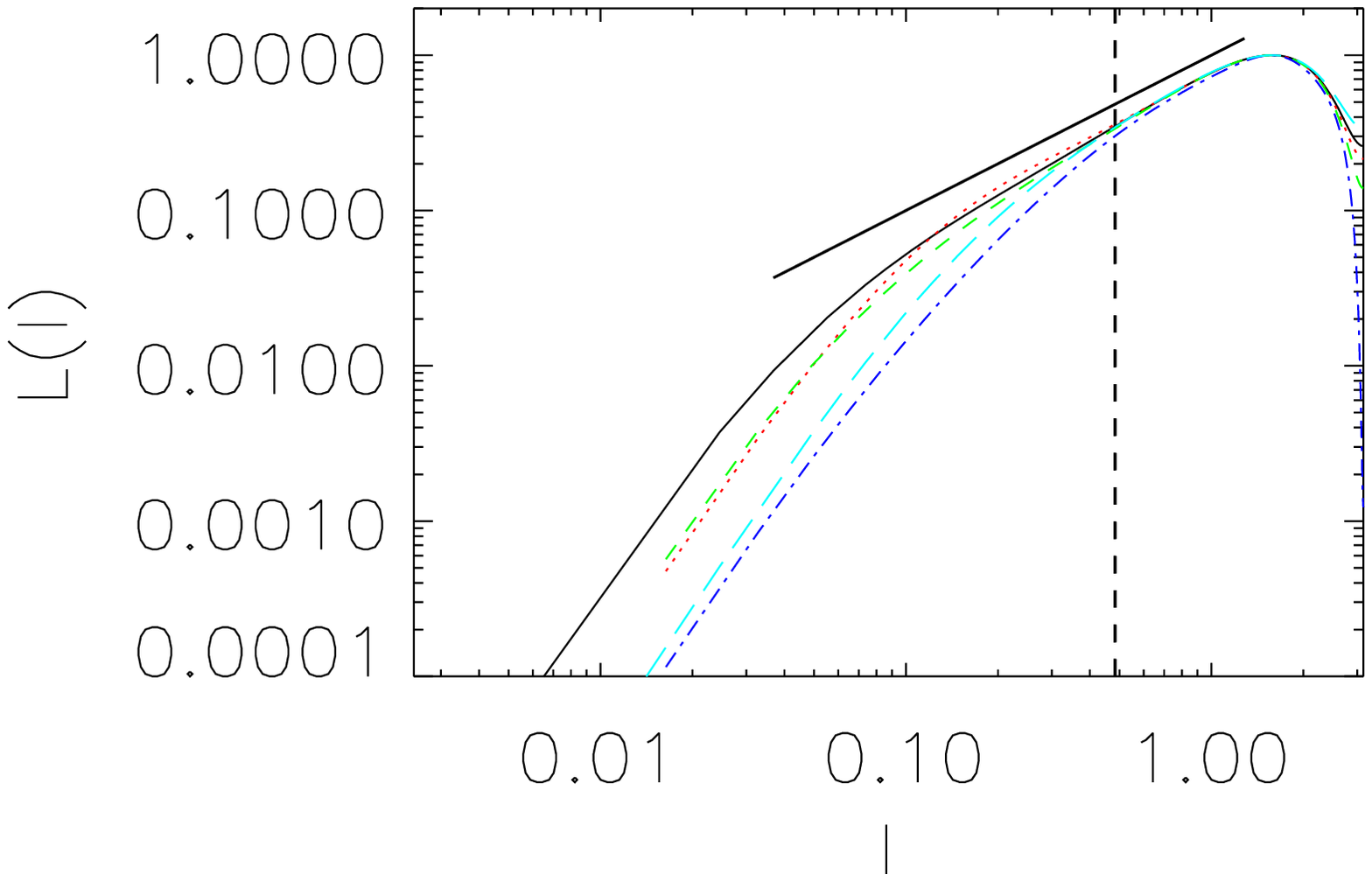} 
  \caption[K\'arm\'an-Howarth scaling] {(Color online) Third-order structure function associated with the
K\'arm\'an-Howarth equation, $L^{(\alpha,L)}(l)$, versus length, $l$.
Labels are as in Fig. \ref{fig:ENSTROPHY}.  The vertical dashed lines
indicate the length $\alpha$. 
\thesis{: $t=9$ for \lansa}
\thesis{: $t=9$ for \lans, $L^\alpha(l) \sim
l^\gamma$ for $k \in [13,21.6]$ with $\gamma = 1.02 \pm 0.04$,
$t\in[9,12.4]$ for \clark, $\gamma = 1.23 \pm 0.05$,
$t=18$ and for \leray, $\gamma = 1.64 \pm 0.16$.}  The \clarka
result is consistent with a $u^2v \sim l^1$ scaling, Eq. (\ref{EQ:LCUBE}), and
clearly inconsistent with a $u^2v \sim l^{-1}$ ($u^3 \sim l$) scaling as would arise from the
middle term in Eq. (\ref{EQ:FULL_CLARK}).
The results for \leraya are again consistent with a reduced effective
$Re$.}
  \label{fig:THEOREM}
\thesis{\end{center}}
\end{figure}
}

Validation of the K\'arm\'an-Howarth equation scalings, Eqs. (\ref{EQ:LCUBE})
and (\ref{EQ:LCUBEleray}), enables us to measure scaling laws in the inertial
range and, thus, compare the intermittency properties of the models.
The third-order correlations involved in the theorems, namely
\begin{equation}
L^\alpha(l) \equiv \left<(\delta u_{\|}(l))^2(\delta v_{\|}(l))\right>
\end{equation}
for \clarka and \lans, and
\begin{equation}
L^L(l) \equiv \left<(\delta v_{\|}(l))^2(\delta u_{\|}(l))\right>
\end{equation}
for \leray, and $L(l) \equiv {S}_3^v(l)$ for Navier-Stokes, are
plotted versus $l$ in Fig. \ref{fig:THEOREM}.
In Fig. \ref{fig:THEOREM} we can see
validation of the K\'arm\'an-Howarth scaling for scales smaller than
$\alpha$ for both \lansa and \clark.  In particular, we note the
observed scaling for \clarka verifies the $vu^2 \sim l$ scaling and
not the (theoretically possible) $vu^2 \sim l^{-1}$ ($u^3 \sim l$) scaling.  The
predicted scaling is not observed in Leray due to its
reduced effective Reynolds number.  With these scalings in hand,
we may proceed to observe the scaling of the longitudinal structure
functions,
\begin{equation}
S^v_p(l) \equiv \langle
({\delta v_\|}^2)^{p/2}\rangle,
\end{equation}
where we again replace the $H^1_\alpha$ norm, $\langle |{\delta
v_\|}||{\delta u_\|}|\rangle$, for the $L^2$ norm, $\langle ({\delta
v_\|})^2\rangle$, in the case of \clarka and \lans.  We utilize the
extended self-similarity (ESS) hypothesis \cite{BCB+93,BCT+93,BBC+96}
which proposes the scaling
\begin{equation}
S_p^v(l) \propto \left(L^{(\alpha,L)}(l)\right)^{\xi_p} \ ,
\nonumber
\end{equation}
and normalize the results by $\xi_3$ to better visualize the deviation
from linearity (which serves as a measure of intermittency).
\add{As we will show in the next section, our flow is anisotropic
in the $z-$direction.  Therefore,
structure functions are computed in horizontal planes only.}  The
results are displayed in Fig. \ref{fig:EXPONENTS}.

\rem{
\begin{figure}[htbp]\thesis{\begin{center}\leavevmode}
    \includegraphics[width=8.95\thesis{11.85}cm]{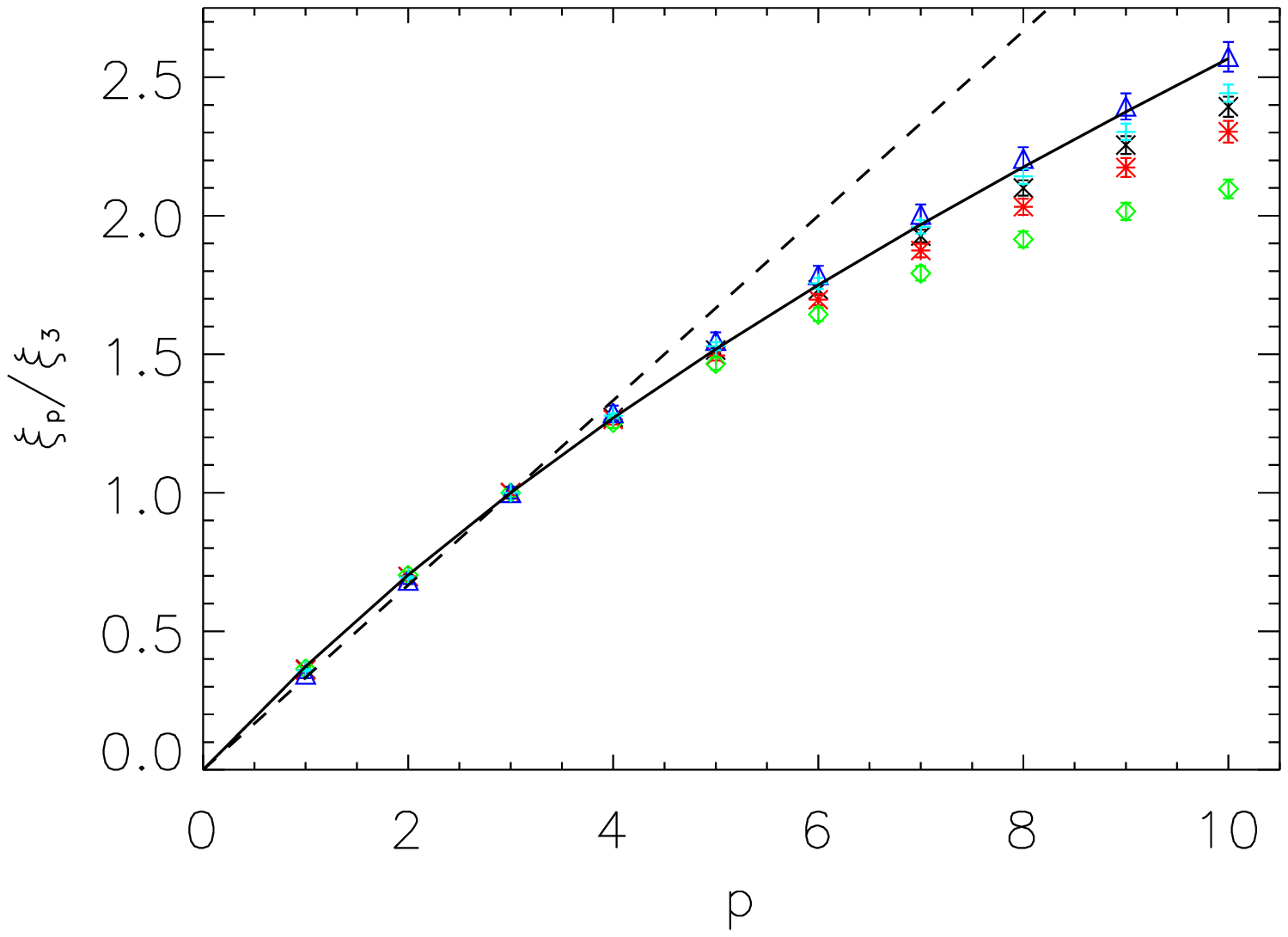} 
  \caption[Normalized structure function scaling exponent versus order] {(Color online) Normalized structure function scaling exponent
$\xi_p/\xi_3$ versus order $p$.  The dashed line indicates K41 scaling
and the solid line the She-L\'ev\^eque formula \cite{SL94}.  The DNS results are
indicated by black X's,  \lansa by
\thesis{$k\in[5,50]$: $t\in[19.1, 21.4]$} red asterisks,
\thesis{$t\in[19.1, 20.3]$} \clarka by green diamonds,
\thesis{$t\in[20.3, 22.5]$}  \leraya by blue triangles, \thesis{$t=9$ for \lansa
$k\in[5,21.6]$, $t\in[9,12.4]$ for \clarka $k\in[5,21.6]$.}  and the
$Re\approx1300$ DNS results are shown by cyan pluses.  \leraya is less
intermittent consistent with the smoother field produced by a lower
$Re$ flow. \clarka is more intermittent than Navier-Stokes at sub-filter scales.
\lansa is less intermittent than
\clark, likely due to the influence of its rigid bodies (see text).}
  \label{fig:EXPONENTS}
\thesis{\end{center}}
\end{figure}
}

In Fig. \ref{fig:EXPONENTS}, we may observe the
intermittency properties of the models at sub-filter scales.  We note
a reduced intermittency for both \leraya and the $Re\approx1300$ DNS.
This is consistent with the smoother, more laminar fields (due to the 
reduction of the 
effective $Re$) possessed by both.  Interestingly, though \lansa and
\clarka both possess the same cascade scaling (Eq. (\ref{EQ:LCUBE}), as
confirmed in Fig. \ref{fig:THEOREM}), the \clarka model is markedly more intermittent than \lans.  If artificially truncated local interactions {(in spectral space) is taken as a cause of} enhanced intermittency
\cite{LDN01,DLN+04}, then the increased intermittency {observed in \clarka} 
is the expected result of truncation of the higher-order terms in the 
sub-filter stress tensor. {Moreover, if the \lans's $\sim k^1$} 
spectrum is indeed associated with rigid bodies, these would serve 
to decrease the intermittency ({no internal degrees of freedom 
being available in} a rigid body) which is consistent with the 
results shown here.  Due to this effect, \lansa of the three
regularization models \add{most resembles} the high-order 
intermittency of \add{Navier-Stokes at sub-filter scales}.  

\section{SGS potential of the regularizations}
\label{sec:SGSSTUDY}

\subsection{Reproduction of super-filter scale properties}
\label{sec:SUPER}

\add{The differences at sub-filter scales between the regularizations
and Navier-Stokes are important to understand how the models may be
improved upon.  From a practical standpoint, an equally important
question is how they predict the super-filter-scale properties of a DNS
when employed as models.  This gives an indication of their SGS modeling
potential.  For this we choose} $\alpha = 2\pi / 40$ corresponding to
an optimal $\alpha-$LES \cite{PGHM+07a}.  Note that the value of
$\alpha$ has been optimized for neither \clarka nor \leray; as a
consequence, these models might perform better in other parameter
regimes than the results indicate in this study.

\begin{figure}[htbp] 
    \includegraphics[width=8.95cm]{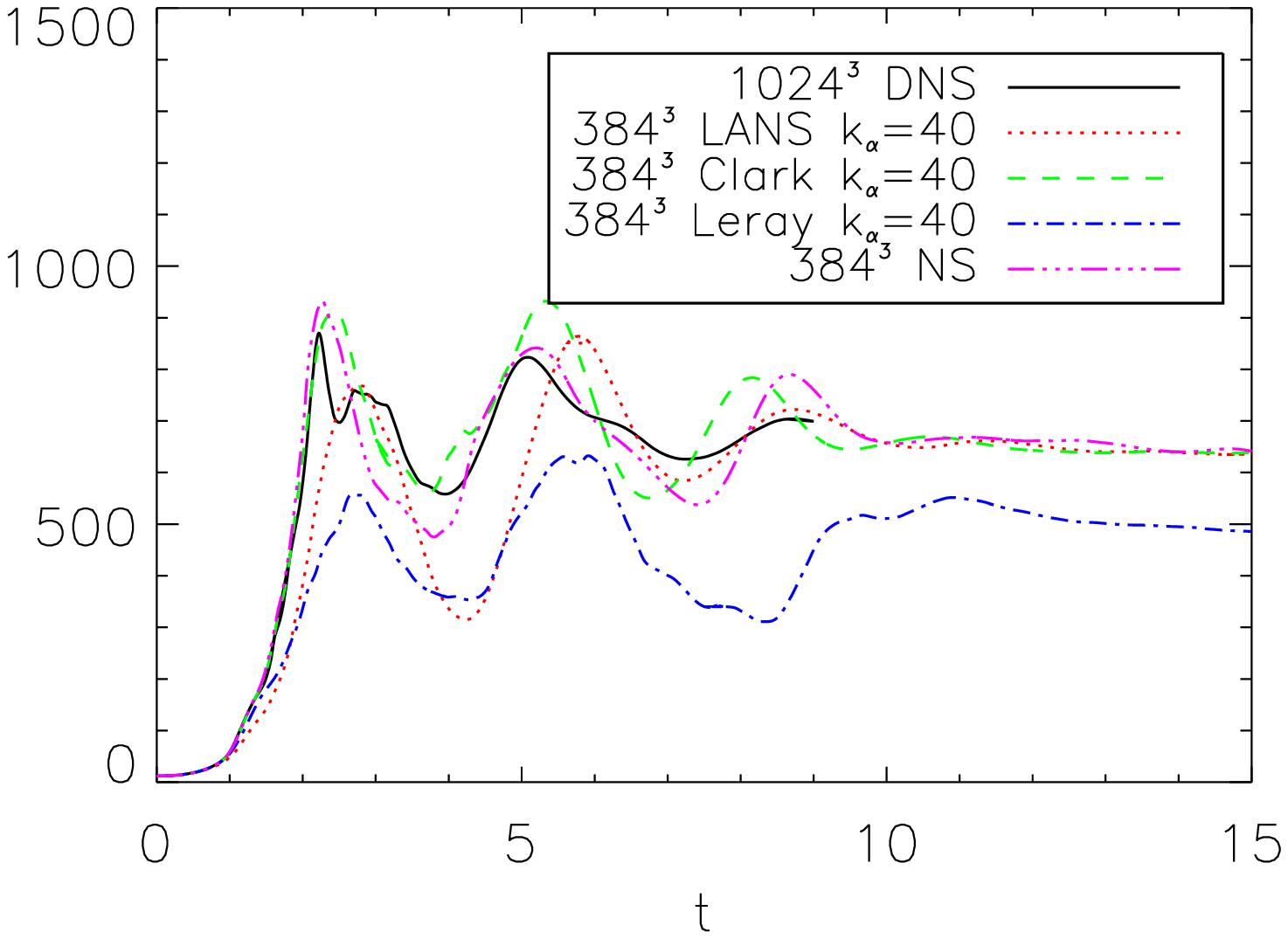} 
  \caption[Time evolution of dissipation] {(Color online) Time evolution of 
  {the enstrophy ($\left<\omega^2\right>$ for \leraya and in the DNS, 
  $\left<\boldsymbol{\omega}\cdot\boldsymbol{\overline{\omega}}\right>$ for 
  \lansa and \clark)}. DNS ($Re\approx3300$) is shown as solid black 
  lines, \lansa as dotted
  red, \clarka as dashed green, and \leraya as blue dash-dotted.
  An under-resolved ($384^3$) Navier-Stokes run
  is shown as a pink dash-triple-dotted line.}
  \label{fig:ENSTROPHY2}
\thesis{\end{center}}
\end{figure}

Fig. \ref{fig:ENSTROPHY2} \add{gives the time evolution of the enstrophy
of the DNS and the models} along with that of an under-resolved
Navier-Stokes solution at a resolution of $384^3$
($\nu=3\cdot10^{-4}$, (pink online) dash-triple-dotted line).  We see
that both \lansa and \clarka reproduce the proper amount of
dissipation and are within $10\%$ of the time required by the DNS to
reach a statistical turbulent steady state.  As has been observed
before, \leraya is under-dissipative \cite{GH06}.  We also note that
it takes longer than the \add{other models to reach a steady state even with
the smaller filter width ($2\pi/40$ as opposed to $2\pi/13$).  When
compared to the larger $\alpha$ case, we see that the dissipation is
much greater and the time-scale to reach a turbulent steady state is
decreased for \leray.}

\rem{
\begin{figure}[htbp]\thesis{\begin{center}\leavevmode}
    \includegraphics[width=8.95\thesis{11}cm]{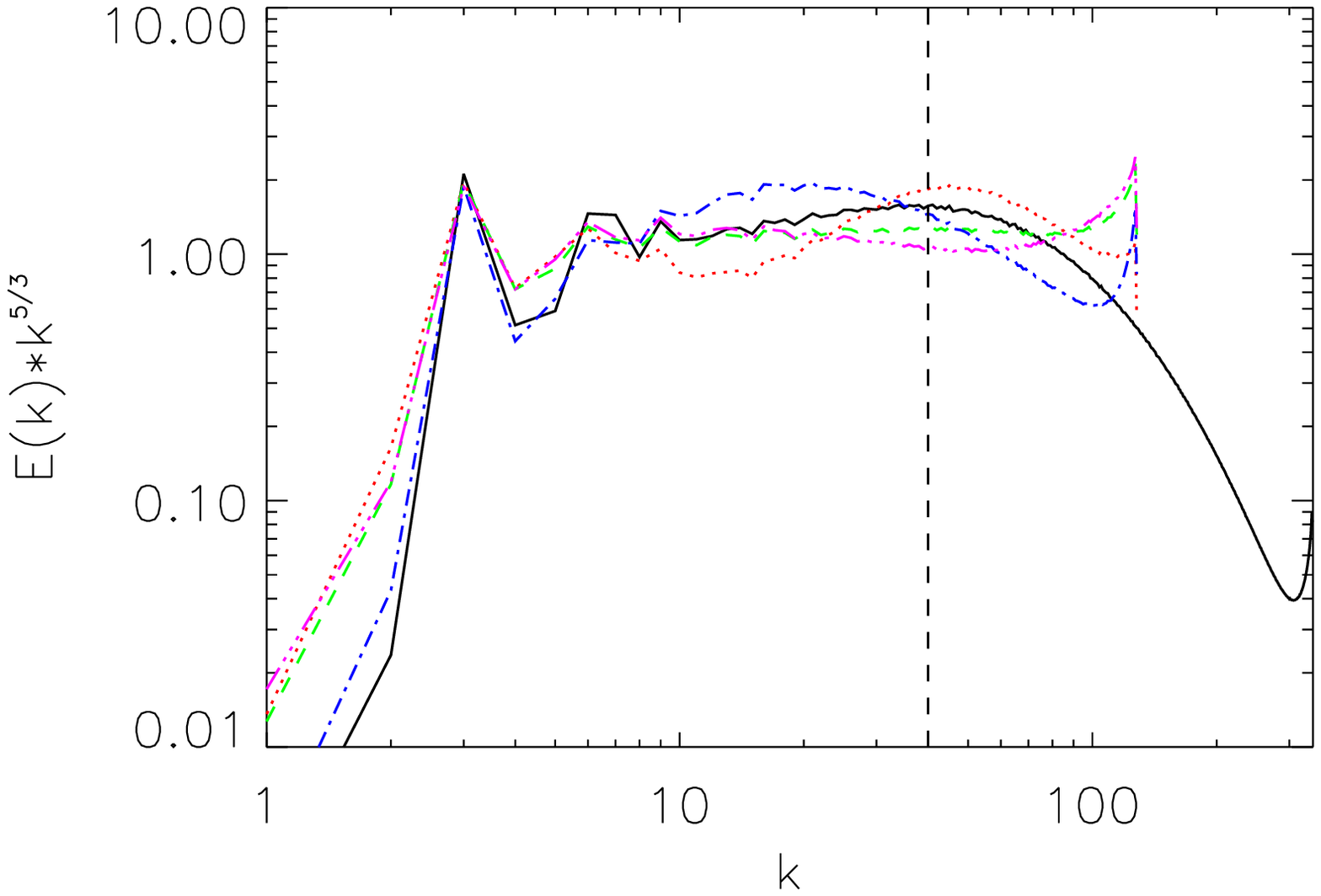} 
  \caption[Averaged compensated spectra] {(Color online) Spectra compensated by K41 for $1024^3$ DNS ($Re\approx3300$) averaged
over $t=[8.25,9]$.  Labels are as in Fig. \ref{fig:ENSTROPHY2}.  \add{The wavenumber
corresponding to the filter width, $k_\alpha$, is shown as a} vertical dashed line.  $384^3$ simulations are
averaged over $t\in[15,20]$.  Note that to make a comparison for most
wavenumbers, the spectra must be averaged within a turbulent
steady-state.  Therefore, as the subgrid models are averaged over a
different time interval, there is no meaningful comparison to the DNS
for $k < 3$. 
\add{For \lansa we observe a contamination of the super-filter-scale
spectrum (at $k\in[9,30]$) related to the steep sub-filter-scale
spectrum and the conservation of energy.}
 Even though a different $\alpha$ \add{(optimized with respect to spectral energy prediction at this numerical resolution) may provide better results} for
\clark, this model does very well at reproducing the large-scale
energy spectrum.  \leray's performance is the poorest.}
  \label{fig:EAVG2}
\thesis{\end{center}}
\end{figure}
}

Compensated spectra averaged over several eddy turn-over times are
shown for the SGS case (i.e., $k_\alpha = 40$) in Fig. \ref{fig:EAVG2}.
Note that as the subgrid models are averaged over a different
time interval, no meaningful comparison to the DNS is possible for $k < k_F = 3$.  Even without an optimal choice for the value of $\alpha$, \clarka
best reproduces the DNS spectrum for scales larger than $\alpha$.  We compute root-mean-square spectral errors \add{as recently introduced in Ref. \cite{MeSaGe2006}:
\begin{equation}
\epsilon_p^b = \left[ \frac{\sum_{k=k_F}^{k_\alpha}k^{2p}(E_{model}(k)-E(k))^2}
{\sum_{k=k_F}^{k_\alpha}k^{2p}E^2(k)} \right]^{1/2},
\label{EQ:ERR_2}
\end{equation}
}where $k_F$ is the wavenumber for the forcing scale, $E(k)$ is the DNS
spectrum (in the $L^2(v)$ norm), and $E_{model}(k)$ is the subgrid model
spectrum (in the appropriate norm).
\add{Another measure introduced in Ref. \cite{MeSaGe2006} is given by
\begin{equation}
\epsilon_p^a = \left[ \frac{\left( \sum_{k=k_F}^{k_\alpha}k^p \left\{E_{model}(k)-E(k)\right\} \right)^2}
{\left( \sum_{k=k_F}^{k_\alpha}k^pE(k) \right)^2}
\right]^{1/2}\,.
\label{EQ:ERR_1}
\end{equation}
With $p=0$, we find the error in the total
energy, $\epsilon_0^a \equiv \epsilon_E$.  As this is dynamically controlled
in our experiment, we find zero in all cases.  For $p=2$, we find the error
in the total dissipation,  $\epsilon_2^a \equiv \epsilon_\varepsilon$,
which is observed in Fig. \ref{fig:ENSTROPHY2}.  Every deviation from the DNS
spectrum is counted positive, however, in $\epsilon_p^b$.
For $p=0$ we find the error in the energy spectrum: in decreasing order, $\epsilon_0^b = 0.24$ for \leray, $0.23$ for
the under-resolved $384^3$, $0.20$ for \lans, and $0.16$ for \clark.
Both \lansa and \clarka improve the estimate over the under-resolved run, but
\clarka makes the best prediction.}
We see that only \clarka improves the estimate of the power spectrum
at this resolution for \add{each scale considered separately (see Fig. \ref{fig:EAVG2}).}  
 \leraya performs the poorest of the three
regularization models, but it is also not optimized.  As
previously argued, its effective Reynolds number is too low to
accurately model the DNS flow.  Either a decrease in the viscosity
$\nu$, or a decrease in the filter size $\alpha$ (and, hence, an
increase in the nonlinearity), or both would likely improve the accuracy of \leray\, as an SGS model.  Due to its frozen-in (or enslaved) rigid-body regions \add{and its conservation of total energy}, the \lansa model cannot reproduce the DNS spectrum \add{at super-filter scales} unless $\alpha$ is only a few
times larger than the dissipation scale \cite{PGHM+07a}.


\rem{
\begin{figure}\thesis{\begin{center}\leavevmode}
\vspace{-1in}
  \begin{tabular}{c@{\hspace{0.02in}}cc}
    \includegraphics[width=4.6cm]{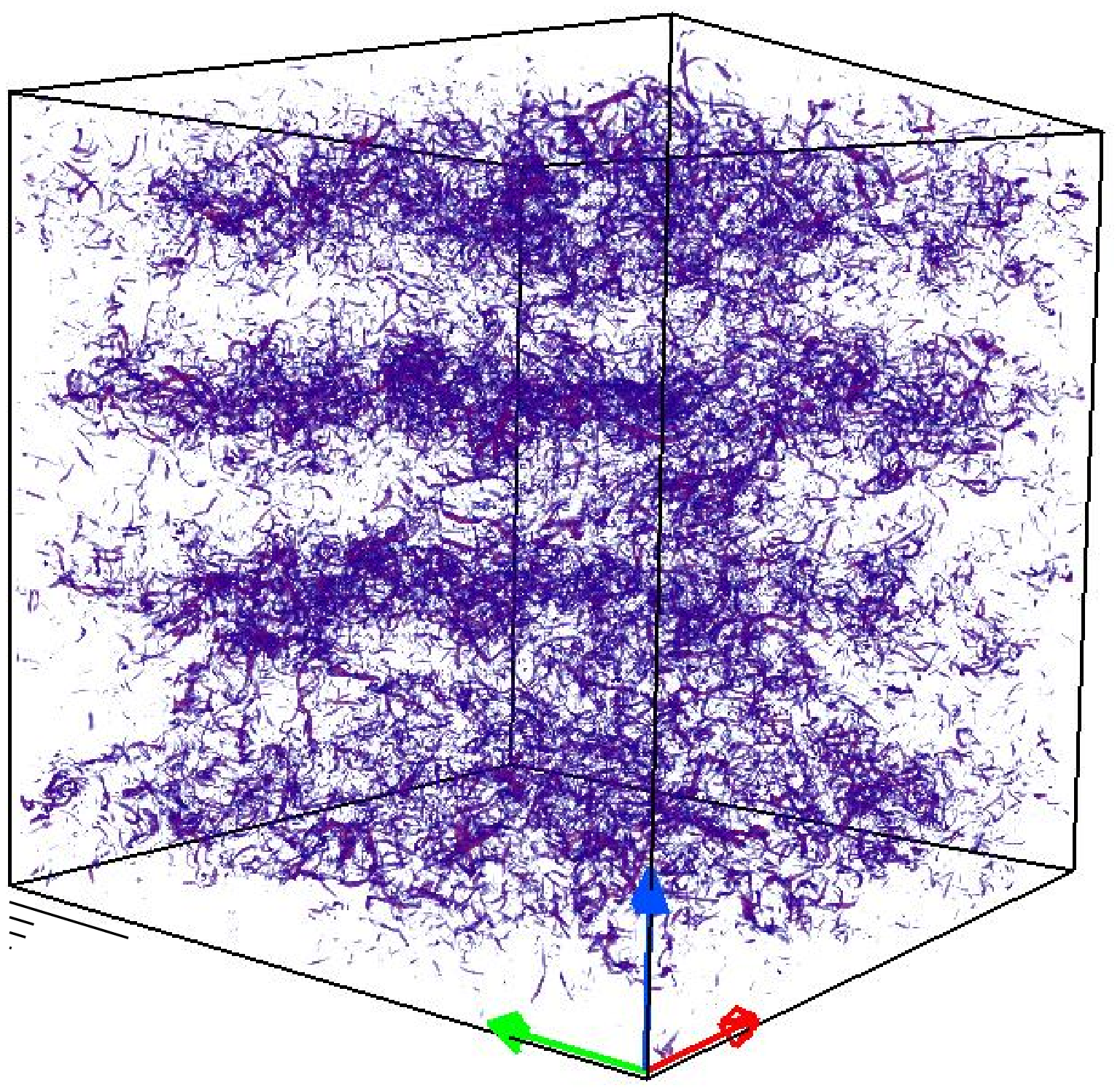} &
    \includegraphics[width=4.6cm]{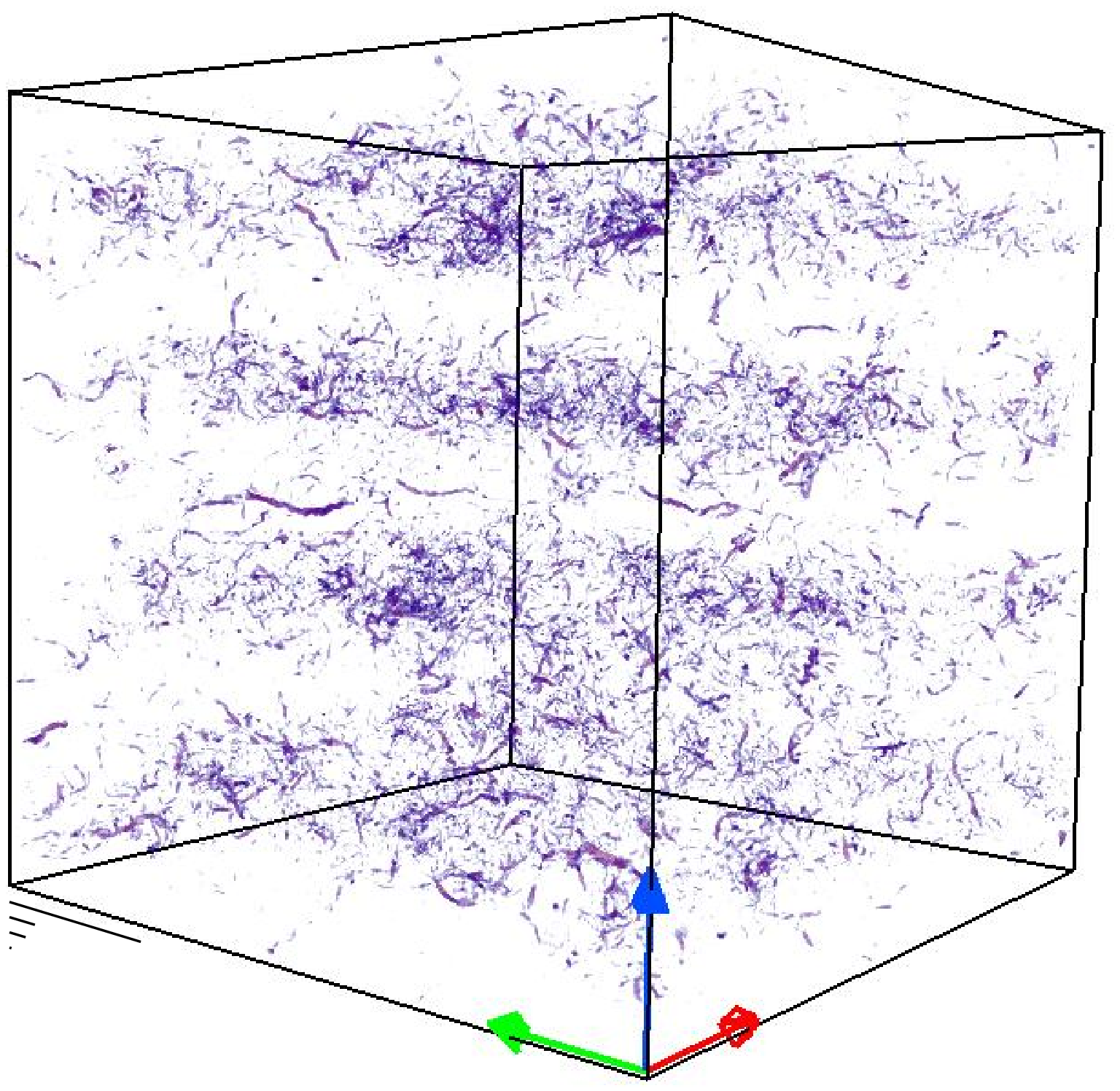} \\
    (a) $Re \approx 3300$ DNS & (b) \clark \\
    \includegraphics[width=4.6cm]{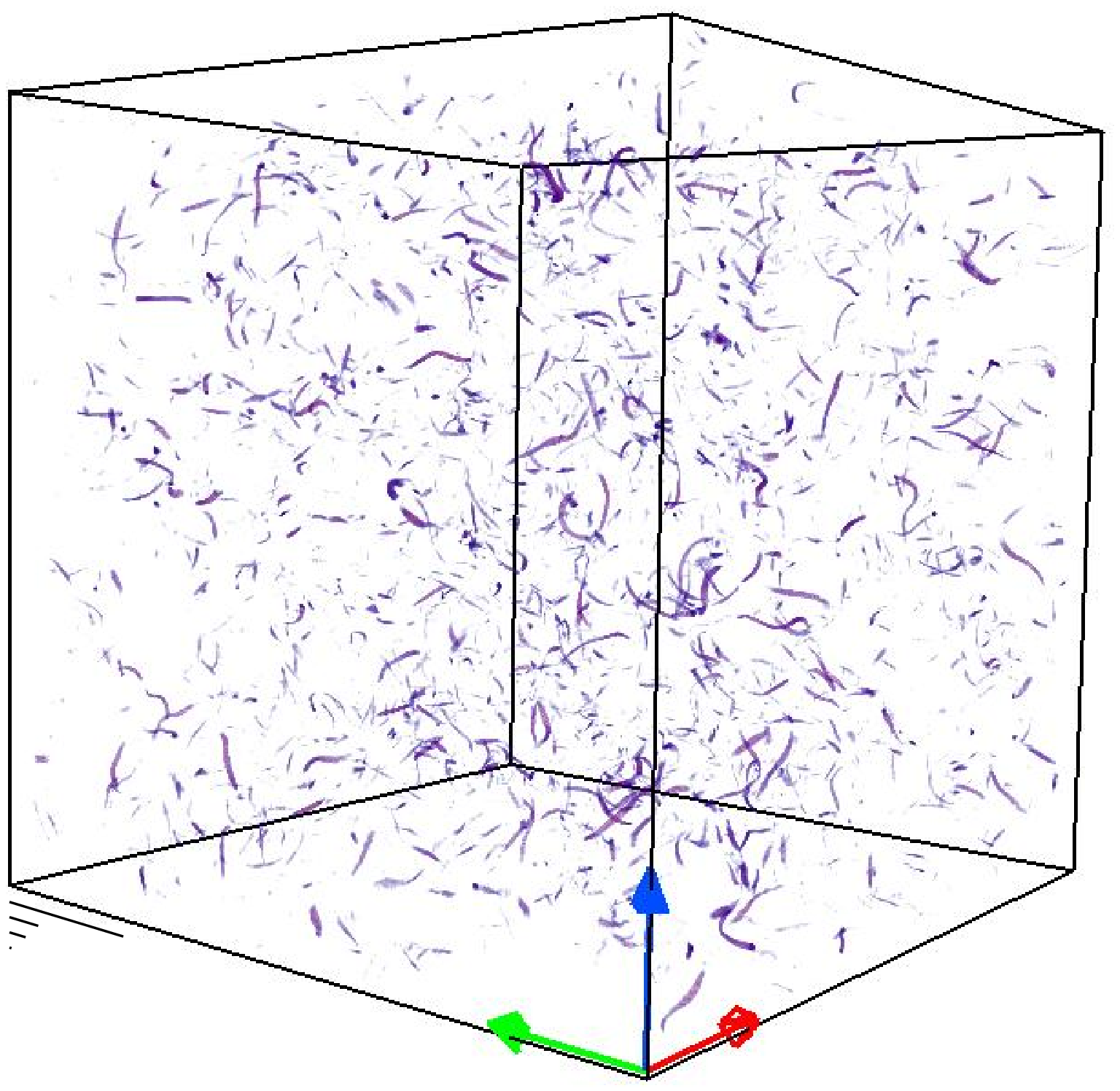} &
    \includegraphics[width=4.6cm]{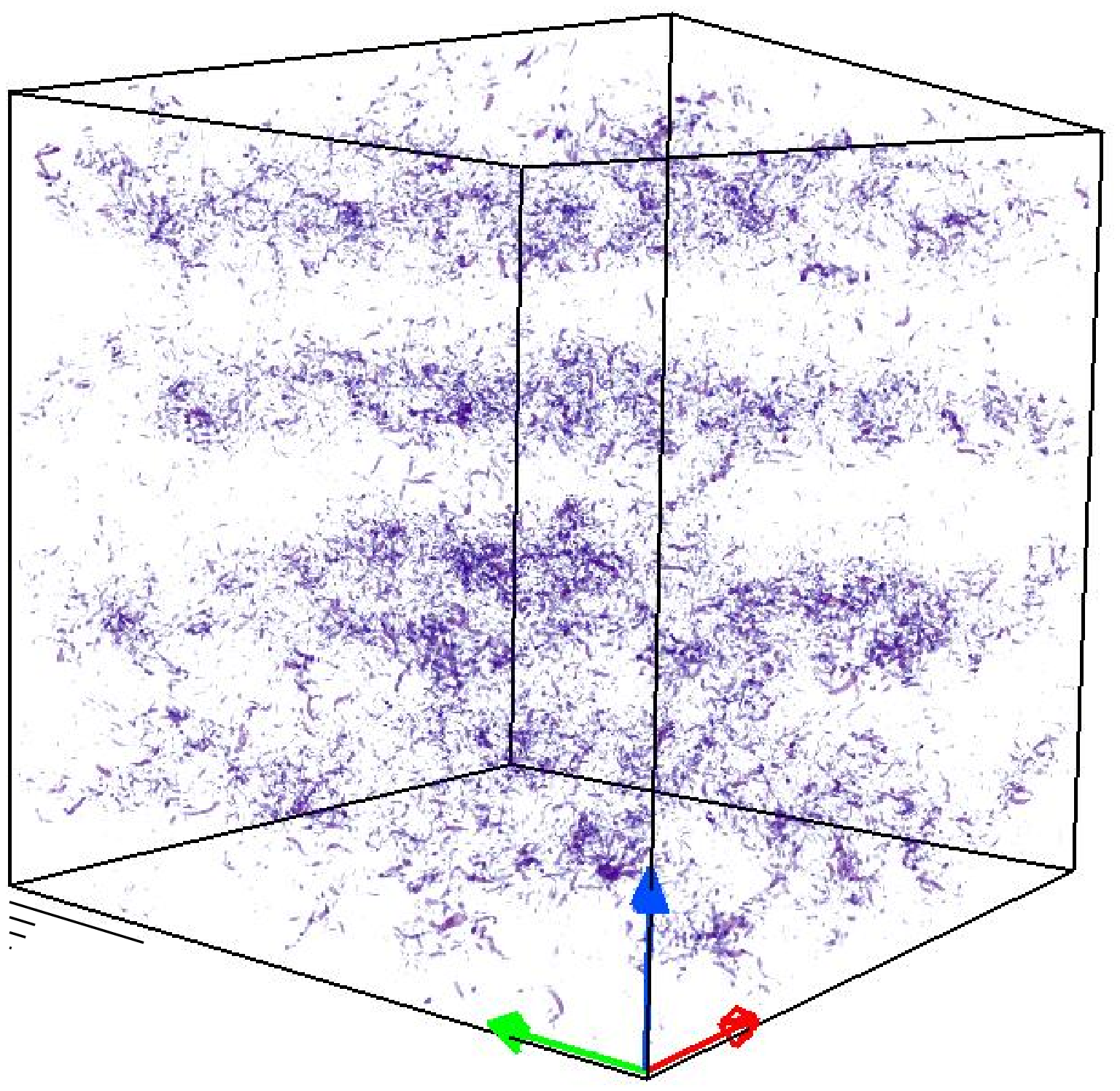} \\
    (c) \leray & (d) \lans \\
    \includegraphics[width=4.6cm]{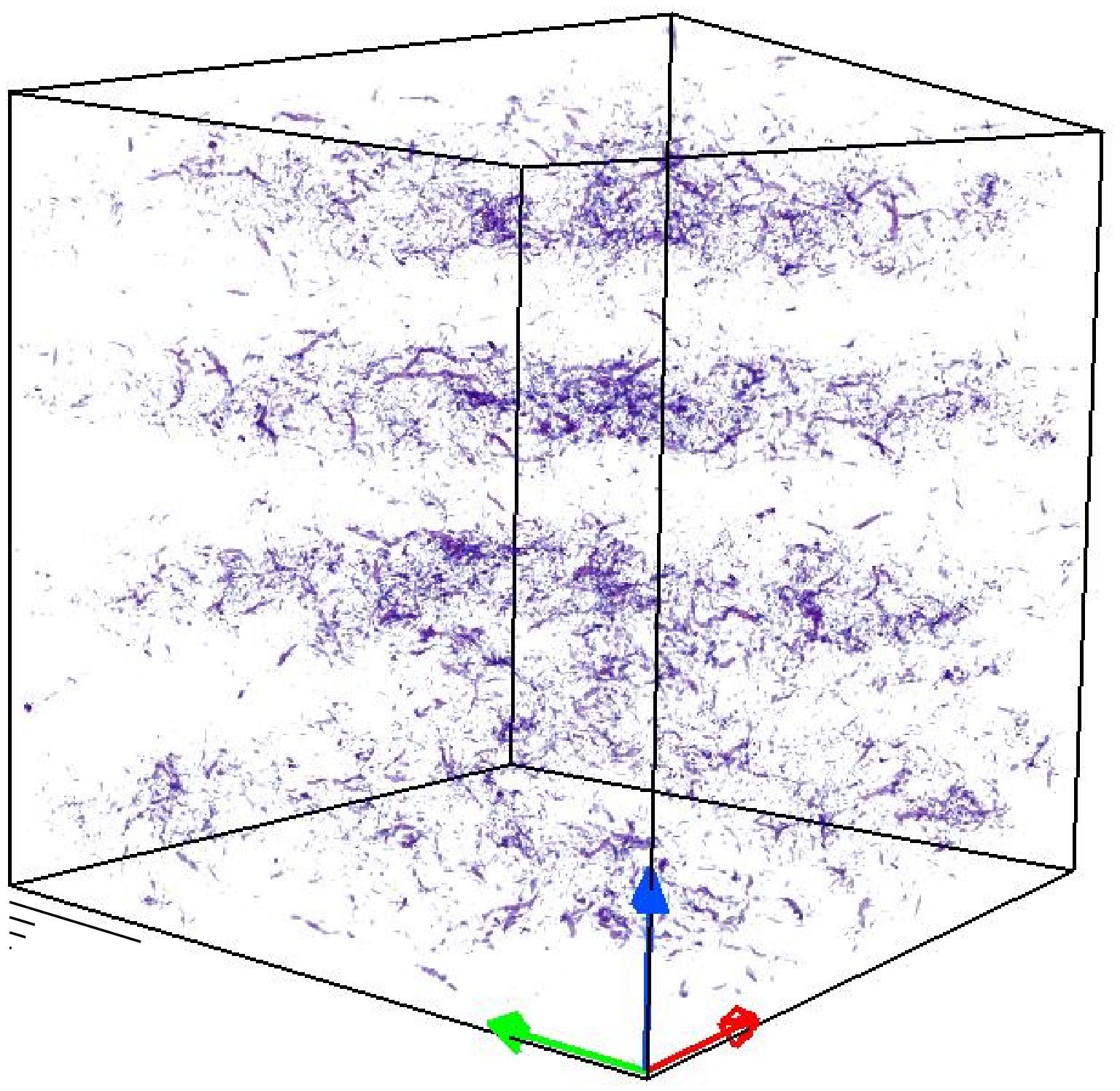} &
    \includegraphics[width=4.6cm]{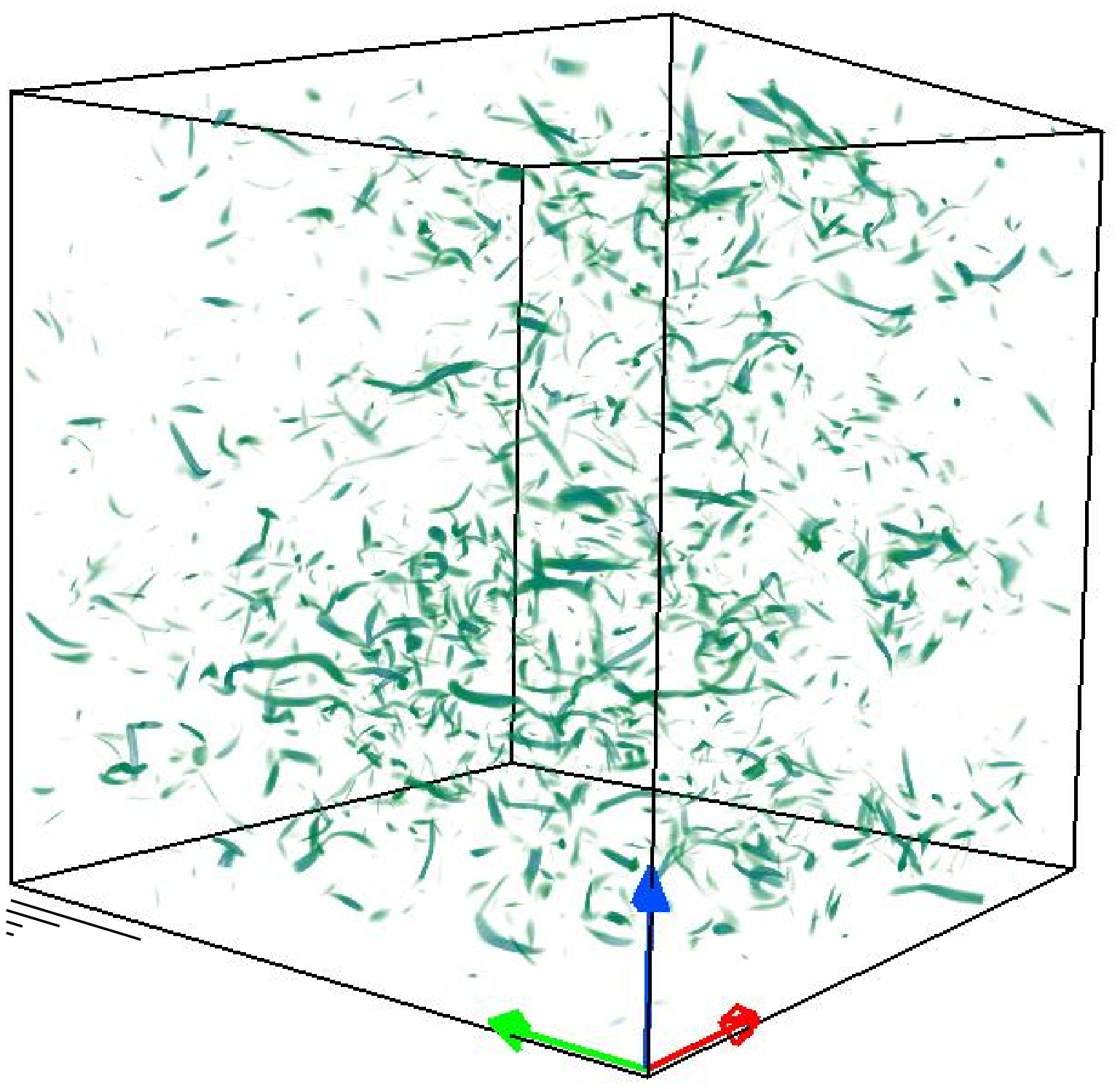} \\
    (e) Under-resolved Navier-Stokes & (f) $Re\approx1300$ DNS \\
  \end{tabular}
  \caption[3D volume rendering of the enstrophy density] {(Color online) Volume
rendering of the enstrophy density $\omega^2$
($\boldsymbol{\omega}\cdot\boldsymbol{\bar\omega}$ for \lansa and \clark). The four
lengths depicted are integral length scale $\mathfrak L$, Taylor scale
$\lambda$, filter width $\alpha$, and dissipative scale $\eta_K$ as
calculated separately for each simulation.  For (a) $Re\approx3300$
DNS, (b) \clark, (d) \lans, and (e) under-resolved Navier-Stokes the
snapshot is for $t=9$.  For (c) \leraya it is for $t=16$ and for (d)
$Re\approx1300$ DNS it for $t=19$ corresponding to their slower
development of turbulence.  For \leraya the location of vortex tubes
are consistent with a lower $Re$ flow while the other models
(including under-resolving) reproduce the large-scale pattern of the
flow well.  The color scale indicates the strength of the enstrophy
density with purple stronger than green.}
  \label{fig:VAPOR}
\thesis{\end{center}}
\end{figure}
}

Another measure of the success of a subgrid model is the reproduction of
structures in the flow.  In Figure \ref{fig:VAPOR} we have 3D volume
rendering of the enstrophy density $\omega^2$
($\boldsymbol{\omega}\cdot\bar{\boldsymbol{\omega}}$ for \lansa and \clark) for the
DNS, the three SGS-model simulations ($k_\alpha=40$), the $384^3$
under-resolved Navier-Stokes solution, all at a Reynolds number of
$\approx3300$, and the $Re\approx1300$ DNS.  Due to the late times
depicted (longer than a Lyapunov time) there can be no point-by-point
comparison between the simulations.  Instead, we note that there are
four horizontal bands where the forcing causes a maximum shear.  This
large-scale feature of the flow is missing only from \leraya and the
$Re\approx1300$ run.  The three other runs reproduce this feature well
(note that the apparently thicker tubes present in \clarka are vortex
tube mergers).  The results lead again to the conclusion that the
under-resolved Navier-Stokes, the \clark, and the \lansa models are {better} subgrid
models than \leraya due to its reduced effective $Re$.


\rem{
\begin{figure}[htbp]\thesis{\begin{center}\leavevmode}
    \includegraphics[width=8.95\thesis{11.85}cm]{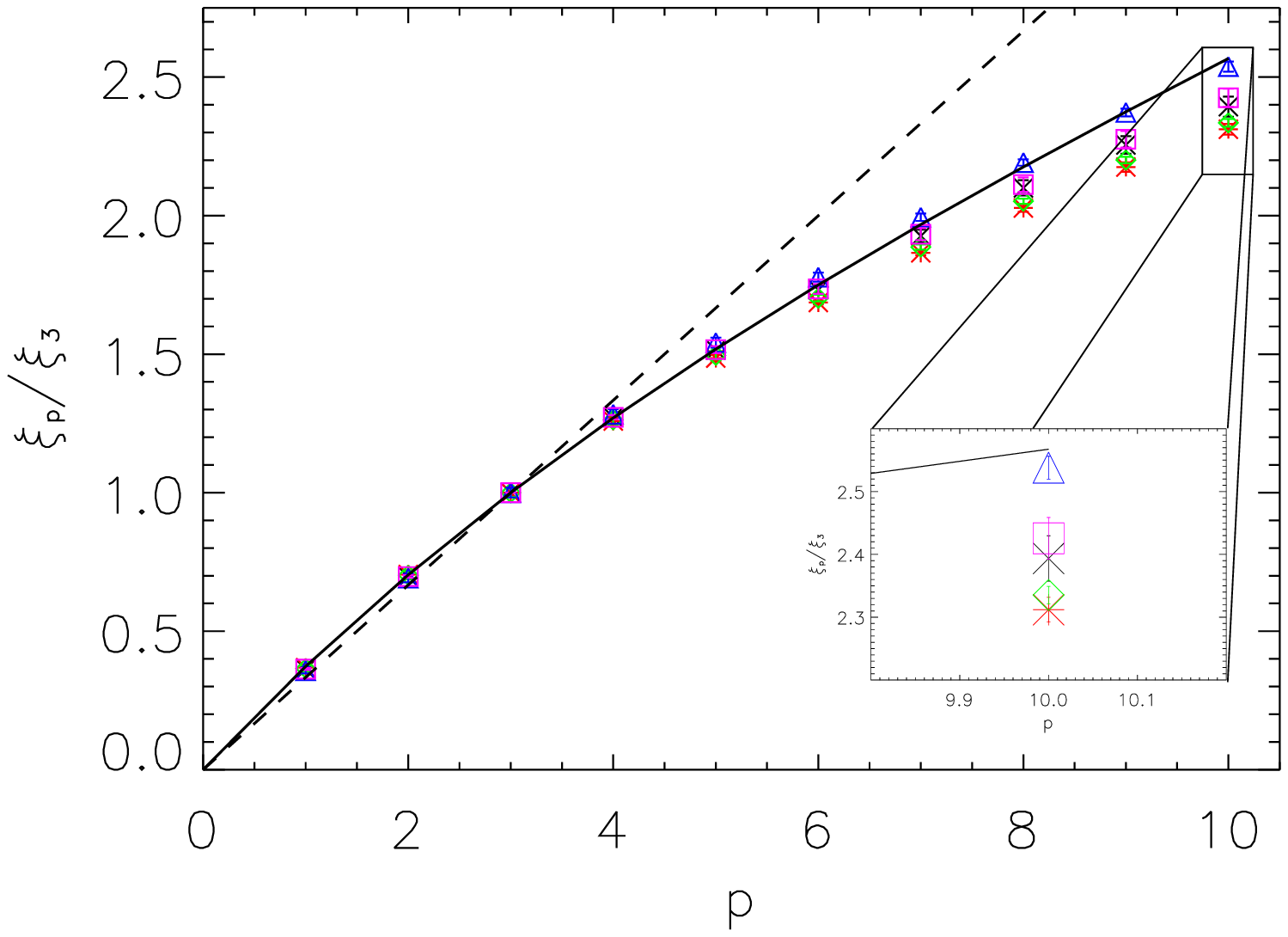} 
  \caption[Normalized structure function scaling exponent versus order] {(Color online) Normalized structure function scaling exponent
$\xi_p/\xi_3$ versus order $p$.  The dashed line indicates K41 scaling
and the solid line the She-L\'ev\^eque formula \cite{SL94}.  The DNS results are
indicated by black X's.
\thesis{$k\in[5,50]$: $t\in[19.1, 21.4]$}  \lansa by red asterisks,
\thesis{$t\in[19.1, 20.3]$}  \clarka by green diamonds,
\thesis{$t\in[20.3, 22.5]$}  \leraya by blue triangles, and
\thesis{$t\in[15.8, 17.9]$} pink boxes for the under-resolved
Navier-Stokes run.  With a small enough filter-width, $\alpha$, the
intermittency properties of the DNS can be reproduced with all three
models. }
  \label{fig:EXPONENTS2}
\thesis{\end{center}}
\end{figure}
}

For the SGS models, the predicted $l^1$ from the K\'arm\'an-Howarth
theorem for Navier-Stokes is well-reproduced by all models at
super-filter scales \add{(not depicted here)}.
\add{We may then proceed}
in Fig. \ref{fig:EXPONENTS2} \add{to analyze}
the SGS model intermittency results.  \add{We see that} all models reproduce the intermittency up
to the tenth-order moment within the error bars (although there is a small decrease in intermittency for \leray).  Thus, we conclude
that with adequately chosen values of $\alpha$ (and of $\nu$ for
\leray), all three models can reproduce the intermittency of the DNS
(to within the error bars).

\add{The sub-filter-scale physics of \leraya shows that it possesses
the smoothest solutions of the three models and reduces the effective
$Re$.  We have seen that this strongly hampers its effectiveness as a
SGS model.  ``Rigid bodies'' are observed in the sub-filter scales of
\lansa \cite{PGHM+07a} that strongly influence even the
super-filter-scale energy spectrum but not the sub-filter-scale
dissipation nor intermittency properties.  These affects also carry
over to its application as a SGS model in that very small filter
widths are required to properly predict the large-scale spectrum.
\clark's approximately $k^{-1}$ sub-filter energy spectrum is the closest to
$k^{-5/3}$ of the three models and is seen to cause the least
contamination of the super-filter-scale spectrum when employed as a
SGS model.  Finally, when the filter width is small enough, the
enhanced intermittency of \clarka is nearly eliminated.}

\subsection{Computational gains}
\label{sec:3DISCUSSION}

The rationale behind using \add{a SGS model} is that it leads to adequate solutions at a reduced computational cost, because it computes fewer  {\sl dof}; indeed, for an \add{SGS model}, the ratio of Navier-Stokes's {\sl dof} to the model's {\sl dof}, a prediction for memory savings and hence computation time
savings for numerical simulation, is a crucial factor.  Consequently, analytical bounds on the sizes of the attractors for the three 
regularization subgrid models may be useful indicators of their computational savings.  The {{\dof} }for \lansa
is derived in \cite{FHT01} and confirmed in {\cite{PGHM+07a},
\begin{equation}
{\dof_\alpha} \propto \frac{ L}{\alpha} Re^{3/2},
\label{eq:dof_alpha}
\end{equation}
where $L$ is the integral scale (or domain size).  We may compare this
to the \dofa for Navier Stokes,
\begin{equation}
\mbox{\sl dof}_{NS} \sim \left(\frac{ L}{\eta_K}\right)^{3} \sim Re^{9/4},
\end{equation}
which immediately yields
\begin{eqnarray}
\frac{{\sl dof}_{NS}}{{\sl dof}_\alpha} \sim (\frac{\alpha}{L})Re^{3/4}.
\label{eq:DOF_lans}
\end{eqnarray}
It was found, however, that to reproduce the \add{super-filter-scale}
energy spectrum of an equivalent DNS, the filter-width $\alpha$ must
be no larger than a few times the dissipation scale, $\eta_K$
{\cite{PGHM+07a}}. This is the result of the ``polymerization'' of the
flow in \lans, and the associated $E(k) \sim k^1$ scaling at sub-filter
scales \add{and consequent contamination at super-filter scales via
energy conservation}. With this added caveat, it follows that the
reduction in \dofa is independent of $Re$ (and a net factor of about
$10$).  Our study here illustrates that the high-order structure
functions may be reproduced for much larger values of $\alpha$.
Therefore, in applications where the spectrum is not of great concern,
much greater reduction in numerical resolution would be feasible.}

For \clarka there is an upper bound on the Hausdorff, $d_H$, and
fractal, $d_F$, dimensions of the attractor,
\begin{equation}
d_H \le d_F \le C \left(\frac{L}{\eta_K^C}\right)^{3}
\left(\frac{L}{\alpha}\right)^{3/4}, 
\end{equation} 
where ${\eta_K^C}$ is the Kolmogorov dissipation length scale
corresponding to the \clarka model \cite{CHT05}.  From its observed
$k^{-1}$ spectrum, we may estimate ${\eta_K^C}$ or, equivalently,
$k_\eta^C \sim 1/{\eta_K^C}$.  For dissipation the large wavenumbers
dominate and, therefore, combining the \clarka energy balance
Eq. (\ref{EQ:LANSA_BALANCE}) with its sub-filter scale energy spectrum
Eq. (\ref{EQ:CLARKA_SPECTRUM}) allows us to implicitly specify its
dissipation wavenumber, $k_\eta^C$, by
\begin{equation}
\frac{\varepsilon_\alpha^C}{\nu} \sim \int^{k_\eta^C} k^2 E_\alpha^C(k)
 dk \sim \int^{k_\eta^C} k^2
 ({\varepsilon_\alpha^C})^{2/3}\alpha^{2/3}k^{-1} dk\\ \sim
({\varepsilon_\alpha^C})^{2/3}\alpha^{2/3}({k_\eta^C})^2.
\end{equation}
Then we have,
\begin{equation}
k_\eta^C \sim \frac{({\varepsilon_\alpha^C})^{1/6}}{\nu^{1/2}\alpha^{1/3}}.
\end{equation}
It follows that
\begin{eqnarray}
\frac{{\sl dof}_{NS}}{{\sl dof}_{Clark}} \sim Re^{3/4}
(\frac{\alpha}{L})^{3/4} \alpha^{-1}.
\label{eq:DOF_clark}
\end{eqnarray}
This is similar to the prediction for \lans, but as energy spectra are
more easily reproduced for larger values of $\alpha$ than with \lansa
(but not the intermittency properties), it may be the case that
$\alpha$ is not tied to the Kolmogorov dissipation scale $\eta_K$.  If
so, then the computational saving might increase as $Re^{3/4}$ which
is promising for use of \clark\,as an LES model.  This conclusion is
bolstered to the extent that the results in Section \ref{sec:SUPER}
for $k_\alpha=40$ ($\alpha \approx 7 \eta_K$) are acceptable.  If
\add{even further separation from the dissipative scale is not
possible}, there is still a greater reduction in \dofa (a factor of
20) for \clark\, than for \lans.

For \leray, we have the following upper bounds on the Hausdorff  dimension, $d_H$, and fractal dimension, $d_F$, of the global attractor,
\begin{equation}
d_H \le d_F \le c \left(\frac{ L}{{\eta_K^L}}\right)^{12/7}
\left(1+\frac{ L}{\alpha}\right)^{9/14}
\end{equation}
where ${\eta_K^L}$ is the dissipation length scale for \leraya
\cite{CHO+05}.  Again, we estimate the dissipation wavenumber for
\leraya $k_\eta^L \sim 1/\eta_K^L$.  Then, from Eqs. (\ref{EQ:LERAY_BALANCE}) and
(\ref{EQ:LERAY_SPECTRUM}), that is, assuming the $k^{-1/3}$ spectrum
resulting from the K\'arm\'an-Howarth equation, we find 
\begin{equation}
\frac{\varepsilon^L}{\nu_L} \sim \int^{k_\eta^L} k^2 E^L(k)
 dk \sim \int^{k_\eta^L} k^2
 ({\varepsilon^L})^{2/3}\alpha^{4/3}k^{-1/3} dk\\ \sim
({\varepsilon^L})^{2/3}\alpha^{4/3}({k_\eta^L})^{8/3}.
\end{equation}
Consequently we have,
\begin{equation}
k_\eta^L \sim \frac{({\varepsilon^L})^{1/8}}{\nu_L^{3/8}\alpha^{1/2}}.
\end{equation}
It follows that
\begin{eqnarray}
\frac{{\sl dof}_{NS}}{{\sl dof}_{Leray}} \sim
\frac{L^{9/7}\nu^{-9/4}}{\nu_L^{-9/14}\alpha^{-6/7}(1 + \frac{L}{\alpha})^{9/14}}.
\label{eq:DOF_leray}
\end{eqnarray}
Our results suggest that for an effective LES the viscosity $\nu_L$ must be chosen to be 
{\sl smaller} than $\nu$.  This leads to an upper bound on the computational
savings for \leray,
\begin{eqnarray}
\frac{{\sl dof}_{NS}}{{\sl dof}_{Leray}} < C
\frac{Re^{45/28}\alpha^{6/7}}{(1 + \frac{L}{\alpha})^{9/14}}.
\end{eqnarray}
If we further assume that $\alpha$ is directly proportional to the dissipative
scale $\eta_K$, we arrive at
\begin{eqnarray}
\frac{{\sl dof}_{NS}}{{\sl dof}_{Leray}} < C Re^{27/56}
\end{eqnarray}
which is not exceedingly promising for use as a LES.  All such estimates
are, however, purely conjectural until the proper choices of $\alpha$ and
$\nu_L$ are determined.

\section{Discussion}
\label{sec:3CONCLUSIONS}

We derived the K\'arm\'an-Howarth equations for the \leraya and
\clarka models.  These two models may be viewed as successive
truncations of the sub-filter scale stress of the Lagrangian-Averaged
Navier-Stokes $\alpha-$model (\lans).  In the case of \clarka two
different inertial range scalings follow from the dimensional analysis
of this equation.  The case of \leraya is simpler as a single scaling
is predicted. This is the case for Navier-Stokes and \lansa as well.
\add{To our knowledge, we computed the first numerical solution of the \clarka model,
the results of which are encouraging for further study.
We compared these to}
solutions for a $1024^3$ DNS under periodic boundary
conditions ($\nu=3\cdot10^{-4}, Re \approx 3300$) using a $384^3$
resolution under the same exact conditions for \lans, \leray, \clark,
and an under-resolved $384^3$ solution of the Navier-Stokes equations.
We employed two different filter widths $\alpha$.  The first choice
$\alpha = 2\pi/13$ was used to \add{understand the sub-filter-scale physics} and
the second choice $\alpha = 2\pi/40$ was employed to test the SGS
potential of the models.  In comparing these two choices, we found for
\leray\, that an increase in $\alpha$ substantially decreases the
nonlinearity (and hence decreases the effective Reynolds number $Re$).
For this reason, we were unable to confirm either the inertial range
scaling from its K\'arm\'an-Howarth equation or its sub-filter scale
energy spectrum.  For \clarka we were able to determine the dominant
K\'arm\'an-Howarth inertial range scaling to be $u^2v\sim l$
\add{which leads to the} associated $k^{-1}$ energy spectrum, \add{also indicated by our results.}

The performance of the three regularizations as SGS models (for a
resolution of $384^3$ and $k_\alpha = 40$) was comparable to that of
the under-resolved Navier-Stokes solution in reproducing the DNS
energy spectrum \add{at super-filter scales}.  Only \clarka showed a
clear improvement in approximating the spectrum.  From 3D volume
rendering of enstrophy density we found that \clarka and \lansa were
comparable to the under-resolved solution.  Even at $\alpha =
2\pi/40$, \leray's 3D spatial structures are consistent with a
significantly reduced $Re$ flow ({e.g., comparable to a} $Re \approx
1300$ {DNS}).  We note that the value of $\alpha$ was chosen optimally
for \lansa at the resolution of $384^3$, and that for \clarka (and
especially for \leray) smaller resolutions \add{(greater computational
savings)} may have comparable results for this value \add{of
$\alpha$}.
Such a comparison is beyond the scope of the present work.

Although \lansa and \clarka exhibit the same inertial range scaling
arising from similarities in their K\'arm\'an-Howarth equations,
\clarka is decidedly more intermittent than \add{Navier-Stokes at sub-filter scales.}  At the same time, \lansa is
only slightly more intermittent than Navier-Stokes.  These results are
consistent with
{the artificial truncation of local nonlinear interactions (in spectral 
space) in the SGS stress tensor of each model}.  This effect is reduced 
for \lansa by the ``rigid-body regions'' enslaved in its larger scale flow which {possess no internal  degrees of freedom}.  
The reduced intermittency observed for \leray\, is related to its smoother, more laminar fields
as a result of its reduced effective $Re$.

Finally, we analyzed the reduction in the number of \dofa in the models, as compared to Navier-Stokes (and, hence, their LES potential {based on their} 
computational savings).  We noted that as \lansa reproduces the intermittency
properties of a DNS quite well even for larger values of $\alpha$, 
{some further} reduction in numerical saving might  
be achieved provided {the contamination due to} its $k^1$ rigid-body energy
spectrum were not important in a given application.  As \clarka
possesses a similar reduction in \dofa to \lans, its LES potential is
tied to the optimal value of $\alpha$ for LES.  Our study indicates
that \clarka may be applicable (especially with regards to the energy
spectrum) for larger values of $\alpha$ than \lans.  In fact, if its
optimal value is not a function of $Re$, the computational resolution
savings increases as $Re^{3/4}$ for \clark.  For the case of
\leray, the prediction is complicated by the effective reduction in $Re$
as $\alpha$ increases.  Prediction of optimized values of $\alpha$ and
of effective dissipation $\nu_L$ are required to assess its LES
potential.  Future work should include such a study for both \leraya
and \clark.

All three regularizations were shown to be successful, in that their
control of the flow gradient reduces the degrees of freedom and saves
computation while preserving a properly defined Reynolds number
(albeit for \leraya that definition is not yet \add{demonstrated}). \clarka
accurately reproduces the total dissipation, the time scale to obtain a
turbulent statistical steady-state and the large-scale energy
spectrum of a DNS.  These results seem to result from \clarka being an order
$\alpha^2$ approximation of Navier-Stokes.  We have shown that \leraya reduces the
effective Reynolds number of the flow.  The last of the three models,
\lansa restores Kelvin's circulation
theorem (advected by a smoothed velocity) and the conservation of a
form of helicity.  Using spectra as a measure of the success of a subgrid model,
\lansa is less than optimal, due to its \add{contamination of the super-filter-scale spectrum.}
However, other measures of the success of a subgrid model are possible: for example, in
regard to intermittency, \lansa may be considered a superior model.
For \clark, intermittency may be a function of filter width while for
\lans, intermittency does not vary much with $\alpha$.

\add{Through examination of these three systems of nonlinear partial differential equations in comparison
to Navier-Stokes, we have demonstrated that intermittency can be
preserved with careful modification of the nonlinearity.  This was
seen with the \lansa model and may be related to the
conservation of small-scale circulation.  Besides intermittency, the
nonlinear terms also play a role in the energy spectrum (at both
sub-filter and super-filter scales) and in the dissipation.  These
terms must model both the nonlocal interactions (to recover the
intermittency) but also the local interactions (which are too strongly
suppressed inside the ``rigid bodies'' of \lans).  Finally, we have
demonstrated that regularization modeling can be employed to reduce
the computational cost while preserving the high-order statistics of
the flow.}

We remark that the computational gain thus far achieved by any of these 
regularizations is insufficient for applications at very high 
Reynolds numbers, and the three subgrid stress tensors 
discussed here may need to be supplemented with an enhanced effective 
viscosity \add{to be employed as LES}. This is a common practice when implementing the Clark model 
(see e.g., Ref. \cite{WWV+01} for a study of this model with an extra 
Smagorinsky term). In this light, the present study may be useful as 
an analysis of the properties of the SGS tensors of the regularizations, 
and to pick best candidates, before the addition of enhanced dissipation.
Studies similar to that in Ref. \cite{PGHM+07a} will also need to be done for the cases of \clarka and \leraya to quantify their computational savings
before the addition of such dissipative terms.



\subsubsection*{Acknowledgements}

Computer time provided at NCAR is gratefully acknowledged, as well as
NSF-CMG grant 0327888.  The work of DDH was partially supported by the
US Department of Energy, Office of Science, Advanced Scientific
Computing Research, and by the Royal Society of London Wolfson Award
for Meritorious Research.


\end{document}